\documentstyle[12pt]{article}
\newcommand{\newsection}{    
\setcounter{equation}{0}
\section}
\def\appendix#1{
  \addtocounter{section}{1}
  \setcounter{equation}{0}
  \renewcommand{\thesection}{\Alph{section}}
  \section*{Appendix \thesection\protect\indent #1}
  \addcontentsline{toc}{section}{Appendix \thesection\ \ \ #1}
  }

\def\e{{\,\rm e}\,}

\def\eop{\vspace*{\fill}\pagebreak}
\def\be{\begin{equation}}
\def\ee{\end{equation}}
\def\bea{\begin{eqnarray}}
\def\eea{\end{eqnarray}}
\def\LA{\left\langle}
\def\RA{\right\rangle}

\newcommand{\rf}[1]{(\ref{#1})}
\newcommand{\eq}[1]{Eq.~(\ref{#1})}

\def\op{operator}
\def\a{\alpha}
\def\b{\beta}
\def\d{\partial}

\def\l{\lambda}
\def\c{\chi}
\def\f{\phi}

\def\om{\omega}

\def\fa{\frac{n!k!}{n_1!n_2!n_3!k_1!k_2!k_3!}}
\def\fb{\frac{n!k!}{n_1!n_2!k_1!k_2!}}
\newcommand{\ie}{{\it i.e.}\ }

\newcommand{\p}{{\prime}}
\newcommand{\ra}{\rightarrow}
\newcommand{\fr}[2]{{\textstyle {#1 \over #2}}}

\newcommand{\non}{\nonumber \\*}

\def\fun#1#2{\lower3.6pt\vbox{\baselineskip0pt\lineskip.9pt
\ialign{$\mathsurround=0pt#1\hfil##\hfil$\crcr#2\crcr\sim\crcr}}}

\begin{document}

\begin{titlepage}
\begin{flushright}
NBI-HE-94-36 \\
hep-ph/9407329 \\
July, 1994
\end{flushright}
\vspace{.5cm}

\begin{center}
{\LARGE Exact Multiparticle Amplitudes at Threshold in} \\
\vspace{0.6cm}{\LARGE $\phi^4$ Theories with Softly
Broken $O(\infty)$ Symmetry}
\end{center} \vspace{1cm}

\begin{center}
{\large Minos\ Axenides}\footnote{E--mail: \ axenides@nbivax.nbi.dk \ }
\\ \mbox{} \\
{\it The Niels Bohr Institute,} \\
{\it Blegdamsvej 17, 2100 Copenhagen, Denmark}\\
\vspace{1cm}
{\large Andrei\ Johansen}\footnote{E--mail: \ ajohansen@nbivax.nbi.dk \ / \
johansen@lnpi.spb.su \ }
\\ \mbox{} \\
{\it St.Petersburg Nuclear Physics Institute,}
\\ {\it Gatchina, 188350 St.Petersburg, Russian Federation}\\
\vspace{1cm}
{\large Yuri\ Makeenko}\footnote{E--mail: \ makeenko@nbivax.nbi.dk \ / \
 makeenko@vxitep.itep.msk.su \ }
\\ \mbox{} \\
{\it The Niels Bohr Institute,} \\
{\it Blegdamsvej 17, 2100 Copenhagen, Denmark} \\ \vskip .2 cm
and  \\  \vskip .2 cm
{\it Institute of Theoretical and Experimental Physics,}
\\ {\it B. Cheremushkinskaya 25, 117259 Moscow, Russian Federation}
\end{center}

\eop

\vspace*{4cm}
\begin{abstract}
We consider the problem of multiparticle production at threshold
in a $\phi^4$-theory with an $O(N_1$$+$$N_2)$ symmetry softly
broken down to $O(N_1)\times O(N_2)$ by nonequal masses.
We derive the set of recurrence relations between the
multiparticle amplitudes which sums all relevant diagrams
with arbitrary number of loops
in the large-$N$ limit with fixed number of produced particles.
We transform it into a quantum mechanical problem and show how it can be
obtained directly from the operator equations of motion by applying the
factorization at large $N$.
We find the exact solutions to the problem
by using the Gelfand--Diki\u{\i} representation of the diagonal
resolvent of the Schr\"{o}dinger operator.
The result coincides with the tree
amplitudes while the effect of loops is the renormalization of the coupling
constant and masses.
The form of the solution is due to the fact that the exact
amplitude of the process $2$$\ra$$n$ vanishes at $n$$>$$2$ on mass shell
when averaged over the $O(N_{1,2})$-indices of incoming particles.
We discuss what dynamical symmetry is behind this property.
We also give an exact solution in the large-$N$ limit for the model of the
$O(N)$$+$$singlet$ scalar particle with the spontaneous breaking of
a reflection symmetry.
\end{abstract}

\vspace{1.5cm}
\noindent
Submitted to {\sl Nuclear Physics B}

\eop
\end{titlepage}
\setcounter{page}{2}

\newsection{Introduction}

The problem of calculating amplitudes of multiparticle production
at threshold has recently received a considerable
interest~\cite{Cor90}--\cite{RS94}.
The explicit results~\cite{Vol92,AKP93a}
for the tree level amplitudes in the $\phi^4$ theory demonstrate the
factorial growth which is expected due to the large number
of identical bosons in the final state.
While the threshold amplitudes
have a vanishing phase volume, they are a lower bound for tree
amplitudes with nonvanishing spatial momenta of produced particles
which violate unitarity~\cite{CT92,AKP92b,Vol92b}.
Therefore,
loops should be taken into account.

The one-loop calculation of the $1$$\ra$$n$ amplitude discovered~\cite{Vol93a}
a very interesting property of the tree $2$$\ra$$n$ threshold amplitude which
is that it vanishes for $n$$>$$4$ when incoming particles are on mass-shell.
This nullification has been extended~\cite{Smi93}--\cite{Smi94}
to more general models and is crucial for
calculations of the amplitude $1$$\ra$$n$ at the one-loop level.  A
dynamical symmetry which may be responsible for the nullification has been
discussed in Ref.~\cite{LRT93b}.
An interesting question is
which properties of the tree and one-loop amplitudes could survive
in the full theory which includes all loop diagrams.

The simplest exactly solvable model is the $O(N)$-symmetric $\phi^4$ theory
in the large-$N$ limit which is solvable~\cite{Wil73} since only the bubble
diagrams contribute if there is no multiparticle production.
In Ref.~\cite{Mak94} the set of recurrence relations
between the amplitudes of
multiparticle production at threshold
has been derived in the large-$N$ limit at fixed number
of produced particles $n$
which sums all relevant diagrams with an arbitrary number of loops.
These recurrence relations have been transformed
into a quantum mechanical problem
whose exact solution has been found through the use of the
Gelfand--Diki\u{\i} representation of the diagonal resolvent
of the Schr\"{o}dinger operator.
The result was quite similar to the tree amplitudes while the effect of
loops is the renormalization of the coupling constant and mass.
The form of the solution was due to the fact that the exact amplitude
of the process $2$$\ra$$n$ vanishes for $n$$>$$2$ at large $N$
on mass shell when averaged over the $O(N)$-indices of
incoming particles.
This nullification occurs for dynamical reasons because of the cancellation
between diagrams.

In the present paper we extend the approach of Ref.~\cite{Mak94}
to more complicated models.
We consider the model of two scalar fields: a $N_1$-component field
with the mass $m_1$ and a $N_2$-component one with the
mass $m_2$. The interaction is $O(N_1$$+$$N_2)$ invariant while
the mass term breaks this symmetry down to $O(N_1)\times O(N_2)$
for nonequal masses $m_1$ and $m_2$. We consider also the case
when the second field is a {\it single\/} component one (\ie $N_2=1$)
and has a wrong sign of the mass square, $m_2^2<0$, which results in the
spontaneous breaking of a reflection symmetry.
The motivations for studying this model are as follows:
\begin{itemize} \vspace{-8pt}
\addtolength{\itemsep}{-8pt}
\item[{\bf 1.}]
The tree amplitudes $1$$\ra$$n$ at threshold can be
found for the model with the softly broken $O(N)$-symmetry extending the
results of Ref.~\cite{LRT93a} given for $N$$=$$2$.
A question arises as to how these
results, obtained exploiting in analogy with $1$-dimensional
integrable systems, can survive at large $N$ in higher loops.
\item[{\bf 2.}]
The model with $N_2$$=$$1$ and $m_2^2<0$ describes after the spontaneous
breaking of a reflection symmetry $1$ particle with mass $\sqrt{2}|m_2|$
and $N$$-$$1$ particles with mass $\sqrt{m_1^2+|m_2^2|}$.
If the $O(N)$-symmetry were not explicitly broken by masses (\ie
if $m_1^2=m_2^2$), these $N$$-$$1$ particles would be massless ---
the Goldstone bosons associated with a spontaneous breaking of
the $O(N)$-symmetry --- so that the consideration would not be possible.
Having the large-$N$ solution, one can perform
a comparison with the one for $N$$=$$1$ which was obtained by semiclassical
methods~\cite{Vol93c,GV93}.
\item[{\bf 3.}]
The model with the softly broken $O(2)$-symmetry reveals at the tree level
a dynamical symmetry which is responsible for the nullification in the
case of further restrictions on kinematics when the mass ratio is an
integer so that no spatial momentum is involved in the decay
$1$$\ra$$n$ \cite{LRT93b}. The problem is whether this dynamical symmetry
exists at arbitrary $m_1$ and $m_2$ even at the tree level.
\vspace{-8pt}
\end{itemize}

We derive the set of recurrence relations between the
multiparticle amplitudes in the large-$N$ limit
at fixed number of produced particles.
We transform it into a quantum mechanical problem and show how it can be
obtained directly from the operator equations of motion by applying the
factorization at large $N$.
We find that
the quantum mechanical problem can be exactly solved through the use of the
Gelfand-Diki\u{\i} techniques.  Thus we get the generating
functions for the amplitudes where the effect of loops is reduced to a
renormalization of the coupling constants and masses.  An interesting new
phenomenon that shows up in the $O(N_1)\times O(N_2)$ model is an appearance of
an imaginary part in the amplitudes which is due to inelastic processes.  An
extension of that result to the case of $O(N_1)\times ...\times O(N_s) ,$ $s>2
,$ is shown to be straightforward.  For the model of the $O(N) +
singlet$ scalar particles, the generating function
for the amplitudes in the large $N$ limit (no loops of a singlet field
should be
taken into account) can be extracted from the explicit solution for the
$O(N_1)\times O(N_2)$ model.

An interesting modification of the result appears when we consider a
model with a negative mass square of the singlet particle.
This model corresponds to a spontaneous breaking of the parity reflection of
the singlet field.
The amplitudes turn out to be real for this model.
By using
the exact solution for this problem we demonstrate a nullification of
amplitudes except for the case of
production of 2 $O(N)$ particles and 1 physical singlet particle.

This paper is organized as follows. Sect.~2 is devoted to
the definitions and the description of the kinematics.  In Sect.~3 we derive
the set of recurrence relations for the $O(N_1)\times O(N_2)$ model and
transform it into a quantum mechanical problem.
In Sect.~4 we present an exact solution and demonstrate that it
satisfies the set of equations.
The exact solution is possible
since the diagonal resolvent of the Schr\"{o}dinger operator
has a very simple form.
In Sect.~5 we consider the $O(N)+singlet$ \/ particle model with
spontaneously broken parity reflection.
We conclude by summarizing the results of this paper and
give a generalization of the dynamical symmetry of
Ref.~\cite{LRT93a} to the $O(N_1)\times O(N_2)$ type models.

\newsection{Definition of the amplitudes}

We consider a model of an $O(N_1)$ scalar field multiplet $\chi^a$ ,
$a=1,...,N_1$ , coupled to a $O(N_2)$ scalar field $\phi^{\alpha} ,$
$\alpha =1,...,N_2 .$
The theory is defined in the 4-dimensional Minkowski space
by the following Lagrangian
\be
{\cal L} = \fr 12 (\partial_\mu \chi^b) (\partial_\mu \chi^b)+
\fr 12 (\partial_\mu \phi^{\a}) (\partial_\mu \phi^{\a})-
\fr 12 m_1^2({\chi}^b{\chi}^b)
-\fr 12 m_2^2(\phi^{\a} \phi^{\a})
-\fr 14 \l (\phi^{\a} \phi^{\a} + {\chi}^b{\chi}^b)^2
\label{lagrangian}
\ee
where the summation over repeated indices is implied.
The interaction term is $O(N_1$$+$$N_2)$ invariant while the mass
term possesses the $O(N_1)$$\times$$O(N_2)$ symmetry for nonequal
masses $m_1$ and $m_2$.

We denote by $A^b_{b_1\ldots b_n,\a_1 \ldots \a_k}
(n,k)$ the amplitude of production of $n$
on-mass-shell particles $\chi$ with the $O(N_1)$-indices
${b_1\ldots b_n}$ and of $k$ on-mass-shell particles $\phi$
with the $O(N_2)$-indices $\a_1, \ldots, \a_k$
at rest by a (virtual) particle $\chi$
with the $O(N_1)$-index $b$ and the energy
$nm_1 +km_2$:
\be
A^b_{b_1\ldots b_n,\a_1 \ldots \a_k}(n,k ) = ((nm_1 +km_2)^2-m_1^2)
m_1^{-n} m_2^{-k}
\LA b_1\ldots b_n,\a_1 \ldots \a_k |\chi^b(0) |0 \RA \,.
\label{defA}
\ee
We also denote by $B^{\b}_{b_1\ldots b_n,\a_1 \ldots \a_k}
(n,k)$ the amplitude of production of $n$
on-mass-shell particles $\chi$ with the $O(N_1)$-indices
${b_1\ldots b_n}$ and of $k$ on-mass-shell particles $\phi$ with
$O(N_2)$-indices $\a_1, \ldots, \a_k$
at rest by a (virtual) particle $\phi$ with the $O(N_2)$-index
$\beta$ and the energy $nm_1 +km_2$:
\be
B^{\b}_{b_1\ldots b_n,\a_1 \ldots \a_k}(n,k ) = ((nm_1 +km_2)^2-m_2^2)
m_1^{-n} m_2^{-k}
\LA b_1\ldots b_n,\a_1 \ldots \a_k | \phi^\beta (0) |0 \RA \,.
\label{defB}
\ee
The extra factors in Eqs.~\rf{defA} and \rf{defB} are due to the
amputation of the external lines.

It is convenient to multiply these amplitudes by
the vectors $\xi_1^{b_i}$ and $\xi_2^{\a_j}$
which results in a symmetrization over the
$O(N_1)$ and $O(N_2)$-indices of the produced particles and define
\be
A^b(n,k) = A^b_{b_1\ldots b_n ,\a_1 \ldots \a_k}(n,k)\,
\xi_1^{b_1}\cdots \xi_1^{b_n} \xi_2^{\a_1} \cdots \xi_2^{\a_k} \, ,
\label{defa}
\ee
and
\be
B^{\a} (n,k) = B^{\a}_{b_1\ldots b_n ,\a_1 \ldots \a_k}
(n,k)\, \xi_1^{b_1}\cdots \xi_1^{b_n}
\xi_2^{\a_1} \cdots \xi_2^{\a_k} \,.
\label{defb}
\ee
The amplitudes $A^b(n,k)$ and $B^{\a}(n,k)$ are
depicted graphically in Fig.~\ref{fig1}$a$,~$b$.
\begin{figure}[tbp]
\unitlength=1.00mm
\linethickness{0.6pt}
\centering
\begin{picture}(118.00,48.00)(-7,80)
\put(15.00,88.00){\makebox(0,0)[cc]{{\Large a)}}}
\put(33.00,114.00){\circle{14.00}}
\put(6.00,114.00){\line(1,0){20.00}}
\put(9.00,112.00){\makebox(0,0)[ct]{$a$}}
\put(33.00,114.00){\makebox(0,0)[cc]{{\large $n,k$}}}
\put(-8.00,114.00){\makebox(0,0)[cc]{{\large $A^a(n,k)=$}}}
\multiput(84.00,114.00)(5.50,0.00){4}{
    \line(1,0){3.00}}
\put(94.00,88.00){\makebox(0,0)[cc]{{\Large b)}}}
\put(111.80,114.00){\circle{14.00}}
\put(88.00,112.00){\makebox(0,0)[ct]{$b$}}
\put(111.80,114.00){\makebox(0,0)[cc]{{\large $n,k$}}}
\put(71.00,114.00){\makebox(0,0)[cc]{{\large $B^\alpha(n,k)=$}}}

\end{picture}
\caption[x]   {\hspace{0.2cm}\parbox[t]{13cm}
{\small
   The graphic representations of the multiparticle
   amplitudes $A^a(n,k)$ (\/{\normalsize a)}\/) and $B^\alpha(n,k)$
   (\/{\normalsize b)}\/). }}
\label{fig1}
\end{figure}

The other quantities which appear in the set of the recurrence
relations between the amplitudes are the amplitudes
$D^{ab}_{b_1\ldots b_n ,\a_1 \ldots \a_k}(n,k;p)$ and
$D^{\a\b}_{b_1\ldots b_n ,\a_1 \ldots \a_k}(n,k;p)$ of the processes
when two particles $a$ and $b$ (or $\a$ and $\b$) with the $4$-momenta
$p+nq_1 +kq_2$ and $-p$, respectively, produce $n$ on-mass-shell particles
$\chi$ with the $O(N_1)$-indices ${b_1,\ldots, b_n}$
and the equal $4$-momenta $q_1 =(m_1 ,0)$ and
$k$ on-mass-shell particles $\phi$ with the $O(N_2)$-indices
${\a_1}, \ldots, {\a_k}$ and
the equal $4$-momenta $q_2 =(m_2 ,0)$ in the rest frame.
We define again
\be
D_{\c}^{ab}(n,k;p)=D^{ab}_{b_1\ldots b_n ,\a_1 \ldots \a_k}
(n,k;p)\, \xi_1^{b_1}\cdots
\xi_1^{b_n}  \xi_2^{\a_1}\cdots \xi_2^{\a_k} \, ,
\label{defD}
\ee
and
\be
D_{\f}^{\a\b}(n,k;p)
=D^{\a\b}_{b_1\ldots b_n ,\a_1 \ldots \a_k}
(n,k;p)\, \xi_1^{b_1}\cdots
\xi_1^{b_n}  \xi_2^{\a_1}\cdots \xi_2^{\a_k} \, .
\label{defP}
\ee
The amplitudes $D_{\c}^{ab}(n,k;p)$ and $D_{\f}^{\a\b}(n,k;p)$
are depicted graphically in Fig.~\ref{fig2}$a$,~$b$.
\begin{figure}[tbp]
\unitlength=1.00mm
\linethickness{0.6pt}
\centering
\begin{picture}(118.00,68.00)(-8,70)
\put(15.00,78.00){\makebox(0,0)[cc]{{\Large a)}}}
\put(30.00,114.00){\oval(14.00,28.00)[]}
\put(30.00,128.00){\line(-1,0){25.00}}
\put(30.00,100.00){\line(-1,0){25.00}}
\put(30.00,114.00){\makebox(0,0)[cc]{{\large $n,k$}}}
\put(16.00,132.00){\makebox(0,0)[cb]{{\large $p$$+$$nq_1$$+$$kq_2$}}}
\put(12.00,96.00){\makebox(0,0)[ct]{{\large $-$$p$}}}
\put(-12.00,114.00){\makebox(0,0)[cc]{
   {\large $D^{ab}_\chi(n,k;p) =$}}}
\put(7.00,126.00){\makebox(0,0)[ct]{$a$}}
\put(7.00,103.00){\makebox(0,0)[cc]{$b$}}

\put(100.00,78.00){\makebox(0,0)[cc]{{\Large b)}}}
\put(115.00,114.00){\oval(14.00,28.00)[]}
\multiput(112.00,128.00)(-6.00,0.00){4}{
    \line(-1,0){3.20}}
\multiput(112.00,100.00)(-6.00,0.00){4}{
    \line(-1,0){3.20}}

\put(115.00,114.00){\makebox(0,0)[cc]{{\large $n,k$}}}
\put(103.00,132.00){\makebox(0,0)[cb]{{\large $p$$+$$nq_1$$+$$kq_2$}}}
\put(99.00,96.00){\makebox(0,0)[ct]{{\large $-$$p$}}}
\put(74.00,114.00){\makebox(0,0)[cc]{
   {\large $D^{\alpha\beta}_\phi(n,k;p) =$}}}
\put(94.00,126.00){\makebox(0,0)[ct]{$a$}}
\put(94.00,103.00){\makebox(0,0)[cc]{$b$}}
\end{picture}
\caption[x]   {\hspace{0.2cm}\parbox[t]{13cm}
{\small
   The graphic representations of the multiparticle
   amplitudes $D^{ab}_\chi(n,k;p)$ (\/{\normalsize a)}\/) and
   $D^{\alpha\beta}_\phi(n,k;p)$ (\/{\normalsize b)}\/). }}
\label{fig2}
\end{figure}
The definitions~\rf{defa}, \rf{defb}, \rf{defD} and~\rf{defP}
generalize those of Ref.~\cite{Mak94} to the case of nonequal
masses $m_1$ and $m_2$.

Notice that we do not introduce the amplitudes associated with
the mixed case when the incoming particles are $\chi$ and $\phi$.
This is
because the unsuppressed in the large $N$ limit loop diagrams
contain only loops of $\chi_b$ and $\f_{\a}$
particles with contracted $O(N_{1,2})$ indices
(it is assumed that $\lambda \sim 1/N$).
Notice also that in the case $N_1 =N \to \infty$ and $N_2=$finite \/ we can
simply drop out the contributions from the loops with $\f$ virtual
particles.
That means that in this case we have to consider only $D^{ab}_{\c}$
vertices.
Actually we shall see
that the amplitudes for the $N_1 =N \to \infty ,$ $N_2=$finite \/ case can
be easily extracted from the expression for the amplitudes given below for
the case $N_{1,2}\to \infty$ by taking formally $N_2/N_1 =0$.

It is easy to estimate at the tree level that
\be
A^b (n,k) \sim \lambda^{\frac{n+k-1}{2}} \left({\xi}_1^2\right)^{\frac{n-1}{2}}
\left(\xi_2^2 \right)^{\frac{k-1}{2}} \xi_1^b ,\;\;\;
B^{\b} (n,k)
\sim  \lambda^{\frac{n+k-1}{2}} \left({\xi}_1^2\right)^{\frac{n-1}{2}}
\left(\xi_2^2 \right)^{\frac{k-1}{2}} \xi_2^{\b} ,
\label{order}
\ee
where $\xi_1^2$ and $\xi_2^2$
stand for $\xi_1^a\xi_1^a$ and $\xi_2^{\a}\xi_2^{\a}$ respectively.
Since
\be
\l \sim \frac 1N
\label{orderl}
\ee
in the large-$N$ limit~\cite{Wil73}, we choose
\be
{\xi_1}^2 \sim {\xi_2}^2 \sim N
\label{orderxi}
\ee
for all the amplitudes~\rf{order} to be of the same order in $1/N$.

\newsection{Schwinger--Dyson equations at large $N$}

\subsection{The diagrammatic derivation}

The condition~\rf{orderl} leaves at large-$N$ only the diagrams with
the largest possible number of sums over internal $O(N_{1,2})$-indices which
propagate along closed loops of diagrams.
The proper recurrence relations
which extend the usual Schwinger--Dyson equations to the case when
$n$ particles $\chi$ and $k$ particles $\phi$ are produced are
depicted for $A^b(n,k)$ and $B^{\a}(n,k)$ in Fig.~\ref{fig3}.
\begin{figure}[tbp]
\unitlength=1.00mm
\linethickness{0.6pt}
\centering
\begin{picture}(108.00,58.00)(41,78)
\put(40.00,114.00){\circle{14.00}}
\put(17.00,114.00){\line(1,0){16.00}}
\put(40.00,114.00){\makebox(0,0)[cc]{{\large $n,k$}}}
\put(58.00,114.00){\makebox(0,0)[cc]{$=$}}
\put(68.00,114.00){\line(1,0){18.00}}
\put(86.00,114.00){\line(1,1){13.30}}
\put(103.00,131.00){\circle{10.00}}
\put(103.00,131.00){\makebox(0,0)[cc]{$n_1,$$k_1$}}
\put(87.00,113.00){\line(1,0){10.90}}
\put(103.00,113.00){\circle{10.00}}
\put(103.00,113.00){\makebox(0,0)[cc]{$n_2,$$k_2$}}
\put(87.00,113.00){\line(1,-1){12.40}}
\put(103.00,97.00){\circle{10.00}}
\put(103.00,97.00){\makebox(0,0)[cc]{$n_3,$$k_3$}}
\put(118.00,114.00){\makebox(0,0)[cc]{$+$}}
\put(128.00,114.00){\line(1,0){18.00}}
\put(146.00,114.00){\line(1,1){13.30}}
\put(163.00,131.00){\circle{10.00}}
\put(163.00,131.00){\makebox(0,0)[cc]{$n_1,$$k_1$}}
\multiput(146.00,113.00)(7.30,0.00){2}{
    \line(1,0){3.60}}

\put(163.00,113.00){\circle{10.00}}
\put(163.00,113.00){\makebox(0,0)[cc]{$n_2,$$k_2$}}
\multiput(146.00,113.00)(8.2,-8.2){2}{
    \line(1,-1){4.00}}
\put(163.00,97.00){\circle{10.00}}
\put(163.00,97.00){\makebox(0,0)[cc]{$n_3,$$k_3$}}
\end{picture}
\begin{picture}(108.00,58.00)(41,80)
\put(61.00,114.00){\makebox(0,0)[cc]{$+$}}
\put(68.00,114.00){\line(1,0){18.00}}
\put(86.00,114.00){\line(1,1){13.30}}
\put(103.00,131.00){\circle{10.00}}
\put(103.00,131.00){\makebox(0,0)[cc]{$n_1,$$k_1$}}
\put(103.00,106.00){\oval(10.00,14.00)[]}
\put(103.00,106.00){\makebox(0,0)[cc]{$n_2,$$k_2$}}
\put(87.00,113.00){\line(1,0){16.00}}
\put(87.00,113.00){\line(1,-1){11.50}}
\put(118.00,114.00){\makebox(0,0)[cc]{$+$}}
\put(128.00,114.00){\line(1,0){18.00}}
\put(146.00,114.00){\line(1,1){13.30}}
\put(163.00,131.00){\circle{10.00}}
\put(163.00,131.00){\makebox(0,0)[cc]{$n_1,$$k_1$}}
\put(163.00,106.00){\oval(10.00,14.00)[]}
\put(163.00,106.00){\makebox(0,0)[cc]{$n_2,$$k_2$}}
\multiput(145.60,113.00)(5.80,0.00){3}{
    \line(1,0){3.00}}
\multiput(145.60,113.00)(8.0,-8.0){2}{
    \line(1,-1){4.00}}
\end{picture}
\caption[x]   {\hspace{0.2cm}\parbox[t]{13cm}
{\small
   The recurrence relation for the amplitude $A^a(n,k)$.
   The analytical formula is given by \eq{recurrencea}.
   The analogous recurrence relation for the amplitude $B^\a(n,k)$
   can be obtained by interchanging the solid and dashed lines.}}
\label{fig3}
\end{figure}
The solid lines are associated
with the propagation of the $O(N_1)$-indices (the field $\chi$),
while the dashed ones stand for the propagation of
the $O(N_2)$ indices (the field $\phi$).
Each vertex is the sum of three possible permutations
of the $O(N)$-indices.  This is taken
into account in Fig.~\ref{fig3} by a combinatorial factor.
The graphic notations becomes clear if one  introduces
the auxiliary field $\sigma(x)=\chi^2(x)+\phi^2(x)$
which propagates in the empty space inside vertices.

The recurrence relations of Fig.~\ref{fig3}
read analytically as follows
\bea
A^a(n,k) = \l \sum_{n_i= odd,\atop k_i=even}
\delta_{n,n_1+n_2+n_3}\delta_{k,k_1+k_2+k_3} \fa \non
\frac{A^a(n_1,k_1)A^b(n_2,k_2)A^b(n_3,k_3)}{((n_1 m_1 +k_1 m_2)^2 -
m_1^2)((n_2 m_1 +k_2 m_2)^2 - m_1^2)((n_3 m_1 +k_3 m_2)^2 - m_1^2)}+\non
+\l \sum_{n_1 ,k_2 ,k_3 = odd,\atop n_2 ,n_3 , k_1=even}
\delta_{n,n_1+n_2+n_3}\delta_{k,k_1+k_2+k_3} \fa \non
\frac{A^a(n_1,k_1)B^{\alpha}(n_2,k_2)B^{\alpha}
(n_3,k_3)}{((n_1 m_1 +k_1 m_2)^2 -
m_1^2)((n_2 m_1 +k_2 m_2)^2 - m_2^2)((n_3 m_1 +k_3 m_2)^2 - m_2^2)}+\non
+\l \sum_{n_1=odd,\atop n_2,k_1,k_2 =even}
\delta_{n,n_1+n_2}\delta_{k,k_1+k_2} \fb \non
\frac{A^a(n_1 ,k_1)}{((n_1m_1 +k_1m_2)^2 - m_1^2)}
\int \frac{d^4 p}{(2\pi)^4}
\left( D_{\chi}^{bb}(n_2,k_2;p) +D_{\phi}^{\alpha\alpha}(n_2,k_2;p)
\right) \,,
\label{recurrencea}
\eea
and similarly
\bea
B^{\alpha} (n,k) = \l  \sum_{k_1 ,n_2 ,n_3= odd,\atop n_1 ,k_2 ,k_3 =even}
\delta_{n,n_1+n_2+n_3}\delta_{k,k_1+k_2+k_3} \fa \non
\frac{B^{\alpha}(n_1,k_1)A^b(n_2,k_2)A^b(n_3,k_3)}{((n_1 m_1 +k_1 m_2)^2 -
m_2^2)((n_2 m_1 +k_2 m_2)^2 - m_1^2)((n_3 m_1 +k_3 m_2)^2 - m_1^2)}+\non
+\l \sum_{k_i= odd, \atop n_i=even}
\delta_{n,n_1+n_2+n_3}\delta_{k,k_1+k_2+k_3} \fa \non
\frac{B^{\alpha}(n_1,k_1)B^{\beta}(n_2,k_2)B^{\beta}
(n_3,k_3)}{((n_1 m_1 +k_1 m_2)^2 -
m_2^2)((n_2 m_1 +k_2 m_2)^2 - m_2^2)((n_3 m_1 +k_3 m_2)^2 - m_2^2)}+\non
+\l \sum_{k_1=odd,\atop n_1,n_2,k_2 =even}
\delta_{n,n_1+n_2}\delta_{k,k_1+k_2} \fb \non
\frac{B^{\alpha}(n_1 ,k_1)}{((n_1m_1 +k_1m_2)^2 - m_2^2)}
\int \frac{d^4 p}{(2\pi)^4}
\left( D_{\chi}^{bb}(n_2,k_2;p) +D_{\phi}^{\alpha\alpha}(n_2,k_2;p)
\right) \,.
\label{recurrenceb}
\eea
Notice that only the summed over the $O(N_1)$ and $O(N_2)$-indices
quantities $D_{\chi}^{bb}(n_2,k_2;p)$ and
$D_{\phi}^{\alpha\alpha}(n_2,k_2;p)$ enter these recurrence relations.

It is the property of large $N_{1,2}$ that the recurrence relations
for $D_{\chi}^{aa}(n,k;p)$ and $D_{\chi}^{aa}(n,k;p)$ express
it via $A^b ,$ $B^{\alpha}$ and $D_{\chi}^{bb} ,$
$D_{\phi}^{\alpha\alpha}$ again.
The recurrence relations for $D_{\chi}^{aa}(n,k;p)$ and
$D_{\phi}^{\alpha\alpha}(n,k;p)$ are depicted in Fig.~\ref{fig4}.
\begin{figure}[tbp]
\unitlength=1.00mm
\linethickness{0.6pt}
\centering
\begin{picture}(108.00,63.00)(41,78)
\put(40.00,114.00){\oval(14.00,28.00)[]}
\put(40.00,128.00){\line(-1,0){22.00}}
\put(40.00,100.00){\line(-1,0){22.00}}
\put(40.00,114.00){\makebox(0,0)[cc]{{\large $n,k$}}}
\put(58.00,114.00){\makebox(0,0)[cc]{$=$}}
\put(68.00,128.00){\line(1,0){18.00}}
\put(87.50,128.00){\line(2,1){11.50}}
\put(103.50,136.00){\circle{10.00}}
\put(103.50,136.00){\makebox(0,0)[cc]{$n_1,$$k_1$}}
\put(87.50,128.00){\line(2,-1){11.50}}
\put(103.50,120.00){\circle{10.00}}
\put(103.50,120.00){\makebox(0,0)[cc]{$n_2,$$k_2$}}
\put(68.00,100.00){\line(1,0){18.00}}
\put(86.00,107.00){\oval(10.00,14.00)[]}
\put(86.00,107.00){\makebox(0,0)[cc]{$n_3,$$k_3$}}
\put(86.00,114.00){\line(0,1){14.00}}
\put(118.00,114.00){\makebox(0,0)[cc]{$+$}}
\put(128.00,128.00){\line(1,0){18.00}}
\multiput(146.00,128.00)(7.4,3.70){2}{
    \line(2,1){4.30}}
\put(163.50,136.00){\circle{10.00}}
\put(163.50,136.00){\makebox(0,0)[cc]{$n_1,$$k_1$}}
\multiput(146.00,128.00)(7.4,-3.70){2}{
    \line(2,-1){4.30}}
\put(163.50,120.00){\circle{10.00}}
\put(163.50,120.00){\makebox(0,0)[cc]{$n_2,$$k_2$}}
\put(128.00,100.00){\line(1,0){18.00}}
\put(146.00,107.00){\oval(10.00,14.00)[]}
\put(146.00,107.00){\makebox(0,0)[cc]{$n_3,$$k_3$}}
\put(146.00,114.00){\line(0,1){14.00}}
\end{picture}
\begin{picture}(108.00,63.00)(41,80)
\put(58.00,114.00){\makebox(0,0)[cc]{$+$}}
\put(68.00,128.00){\line(1,0){18.00}}
\put(102.50,128.00){\oval(10.00,14.00)[]}
\put(102.50,128.00){\makebox(0,0)[cc]{$n_1,$$k_1$}}
\multiput(146.00,128.00)(7.6,-3.80){2}{
    \line(2,-1){4.40}}
\multiput(146.00,128.00)(7.6,3.80){2}{
    \line(2,1){4.40}}
\put(87.50,128.00){\line(2,1){11.50}}
\put(87.50,128.00){\line(2,-1){11.50}}
\put(68.00,100.00){\line(1,0){18.00}}
\put(86.00,107.00){\oval(10.00,14.00)[]}
\put(86.00,107.00){\makebox(0,0)[cc]{$n_2,$$k_2$}}
\put(86.00,114.00){\line(0,1){14.00}}
\put(121.00,114.00){\makebox(0,0)[cc]{$+$}}
\put(128.00,128.00){\line(1,0){18.00}}
\put(162.50,128.00){\oval(10.00,14.00)[]}
\put(162.50,128.00){\makebox(0,0)[cc]{$n_1,$$k_1$}}
\multiput(146.00,128.00)(7.6,-3.80){2}{
    \line(2,-1){4.40}}
\multiput(146.00,128.00)(7.6,3.80){2}{
    \line(2,1){4.40}}
\put(128.00,100.00){\line(1,0){18.00}}
\put(146.00,107.00){\oval(10.00,14.00)[]}
\put(146.00,107.00){\makebox(0,0)[cc]{$n_2,$$k_2$}}
\put(146.00,114.00){\line(0,1){14.00}}
\end{picture}
\caption[x]   {\hspace{0.2cm}\parbox[t]{13cm}
{\small
   The recurrence relation for the amplitude $D_\chi^{aa}(n,k;p)$.
   The analytical formula is given by Eq.~\rf{recurrenceD}.
   The recurrence relation for the amplitude $D_\phi^{\a\a}(n,k;p)$
   can be obtained by interchanging the solid and dashed lines. }}
\label{fig4}
\end{figure}

The recurrence relation of Fig.~\ref{fig4}
reads analytically as follows%
\footnote{Here and below the poles should be understood
according to the Feynman prescription $m^2\ra m^2-i0$.}
\bea
D_{\chi}^{aa}(n,k) = \l \sum_{n_1,n_2=odd, \atop n_3,k_i = even}
\delta_{n,n_1+n_2+n_3}\delta_{k,k_1+k_2+k_3} \fa \non
\frac{1}{(p+n_3q_1 +k_3q_2)^2-m_1^2}
\frac{A^b(n_1,k_1)A^b(n_2,k_2)D_{\chi}^{aa}(n_3,k_3;p)}{((n_1m_1 +k_1m_2)^2
- m_1^2)((n_2m_1 +k_2m_2)^2 - m_1^2)} +\non
+\l \sum_{k_1,k_2=odd, \atop n_i,k_3 = even}
\delta_{n,n_1+n_2+n_3}\delta_{k,k_1+k_2+k_3} \fa \non
\frac{1}{(p+n_3q_1 +k_3q_2)^2-m_1^2}
\frac{B^{\beta}(n_1,k_1)B^{\beta}
(n_2,k_2)D_{\chi}^{aa}(n_3,k_3;p)}{((n_1m_1 +k_1m_2)^2
- m_2^2)((n_2m_1 +k_2m_2)^2 - m_2^2)}+\non
+\l \sum_{n_i,k_i=even}
\delta_{n,n_1+n_2}\delta_{k,k_1+k_2} \fb \non
\frac{1}{(p+n_2q_1+k_2q_2)^2-m_1^2}
\int \frac{d^4 k}{(2\pi)^4} (D_{\chi}^{bb}(n_1,k_1;k)+
D_{\phi}^{\beta\beta}(n_1,k_1;k) ) D^{aa}_\chi(n_2,k_2;p) \,.
\label{recurrenceD}
\eea
Similarly we have
\bea
D_{\phi}^{\a\a}(n,k) = \l \sum_{n_1,n_2=odd, \atop n_3,k_i = even}
\delta_{n,n_1+n_2+n_3}\delta_{k,k_1+k_2+k_3} \fa \non
\frac{1}{(p+n_3q_1 +k_3q_2)^2-m_2^2}
\frac{A^b(n_1,k_1)A^b(n_2,k_2)D_{\phi}^{\a\a}(n_3,k_3;p)}{((n_1m_1 +k_1m_2)^2
- m_1^2)((n_2m_1 +k_2m_2)^2 - m_1^2)} +\non
+\l \sum_{k_1,k_2=odd, \atop k_3,n_i = even}
\delta_{n,n_1+n_2+n_3}\delta_{k,k_1+k_2+k_3} \fa \non
\frac{1}{(p+n_3q_1 +k_3q_2)^2-m_2^2}
\frac{B^{\beta}(n_1,k_1)B^{\beta}
(n_2,k_2)D_{\phi}^{\a\a}(n_3,k_3;p)}{((n_1m_1 +k_1m_2)^2
- m_2^2)((n_2m_1 +k_2m_2)^2 - m_2^2)}+ \non
+\l \sum_{n_i,k_i=even}
\delta_{n,n_1+n_2}\delta_{k,k_1+k_2} \fb \non
\frac{1}{(p+n_2q_1+k_2q_2)^2-m_2^2}
\int \frac{d^4 k}{(2\pi)^4} (D_{\chi}^{bb}(n_1,k_1;k)+
D_{\phi}^{\beta\beta}(n_1,k_1;k) ) D^{\a\a}_\phi(n_2,k_2;p) \,.
\label{recurrenceG}
\eea
Eqs.~\rf{recurrencea}, ~\rf{recurrenceb}, \rf{recurrenceD}
and \rf{recurrenceG}
look very similar to the ones~\cite{Mak94} for
the $O(N)$-case.

To rewrite Eqs.~\rf{recurrencea}, ~\rf{recurrenceb}, \rf{recurrenceD}
and \rf{recurrenceG} in a more
convenient form, let us introduce
the generating functions
\be
\Phi^a(\tau) = m_1 \xi_1^a \e^{m_1\tau}  + \sum_{n,k\geq3}
\frac{A^a(n,k)}{n!k!((nm_1 +km_2)^2 - m_1^2)} \e^{(nm_1+km_2)\tau} m_1^n
m_2^k ,
\label{Phi}
\ee
\be
\Psi^{\a}(\tau) = m_2 \xi_2^{\a} \e^{m_2\tau}  + \sum_{n,k\geq3}
\frac{B^{\a}(n,k)}{n!k!((nm_1 +km_2)^2 - m_2^2)} \e^{(nm_1+km_2)\tau}
m_2^n m_2^k ,
\label{Psi}
\ee
\be
D_{\c}^{ab}(\tau;p) = \frac{i\,\delta^{ab}}{p^2-m_1^2}  + \sum_{n,k=even}
D_{\c}^{ab}(n,k;p) \frac{1}{n!k!} \e^{(nm_1+km_2)\tau} m_1^{n} m_2^k
\label{D}
\ee
and
\be
D_{\f}^{\a\b}(\tau;p) = \frac{i\,\delta^{\a\b}}{p^2-m_2^2}  + \sum_{n,k=even}
D_{\f}^{\a\b}(n,k;p) \frac{1}{n!k!} \e^{(nm_1+km_2)\tau} m_1^{n} m_2^k \,.
\label{Dphi}
\ee
Eqs.~\rf{recurrencea}, ~\rf{recurrenceb}, \rf{recurrenceD}
and \rf{recurrenceG} can then be rewritten,
respectively, as
\be
\left\{\frac{d^2}{d\tau^2} -m_1^2- v(\tau)\right \} \Phi^a(\tau) =0,
\label{eq1}
\ee
\be
\left\{\frac{d^2}{d\tau^2} -m_2^2- v(\tau)\right \} \Psi^{\a}(\tau) =0
\label{eqf}
\ee
and
\be
\left\{\frac{d^2}{d\tau^2} -\om_1^2- v(\tau)  \right\}
\e^{\epsilon \tau}\frac 1{N_1} D_{\c}^{bb}(\tau;p)=
\e^{\epsilon \tau} \,,
\label{eq2}
\ee
\be
\left\{\frac{d^2}{d\tau^2} -\om_2^2- v(\tau)  \right\}
\e^{\epsilon \tau}\frac 1{N_2} D_{\f}^{bb}(\tau;p)=
\e^{\epsilon \tau}
\label{eqff}
\ee
where $\epsilon$ is the energy component of $p$,
$p=(\epsilon,\vec{p}\,)$,
\be
\om_{1,2} = \sqrt{\vec{p}\;{}^2+m_{1,2}^2}
\label{om12}
\ee
and
\be
v(\tau) = \l (\Phi^2(\tau)+\Psi^2(\tau)) + \l \int \frac{d^4
p}{(2\pi)^4} (D_{\c}^{bb}(\tau;p)+D_{\f}^{\b\b}(\tau;p)) \,.
\label{u}
\ee

The next step in the transformation of
Eqs.~\rf{eq1}, \rf{eqf}, \rf{eq2}, \rf{eqff}
and \rf{u} is to put $D_\chi^{ab}$ and $D_\phi^{\a\b}$ in the mixed
representation --- the coordinate in (imaginary) time and
momentum for space.
One defines
\be
D_{\c}^{ab}(\om_1 ;\tau,\tau^\prime) = \int \frac{d\epsilon}{2\pi}
\e^{\epsilon(\tau-\tau^\prime)} D_{\c}^{ab}(\tau;p) \,,
\label{Fourierc}
\ee
\be
D_{\f}^{\a\b}(\om_2 ;\tau,\tau^\prime) = \int \frac{d\epsilon}{2\pi}
\e^{\epsilon(\tau-\tau^\prime)} D_{\f}^{\a\b}(\tau;p) \,.
\label{Fourierf}
\ee
In order to see that this quantity is indeed associated with the
Fourier transform of  amplitudes in energy, we notice that the
insertion of \eq{D} on the r.h.s.\ of \eq{Fourierc} yields
\be
D^{ab}_\c(\om_1 ;\tau,\tau^\prime) = \int \frac{d\epsilon}{2\pi}
\sum_{n=0}^\infty
\e^{(\epsilon+nm_1 +km_2)\tau-\epsilon\tau^\p}\; D_{\c}^{ab}(n;p) \,.
\label{Fourier1}
\ee
One recognizes now that
$\epsilon+nm_1 +km_2$ is the energy component of the
$4$-momentum $p+nq_1 +kq_2$ of the incoming particle with the
$O(N_1)$-index $a$ while
$\epsilon$ is that of $p$ for $b$. In particular, the free propagator in the
mixed representation is given by
\be
D_{\c}^{ab} (\om_1 ;\tau,\tau^\p)=i\,\delta^{ab}
\int \frac{d\epsilon}{2\pi}
\frac{\e^{\epsilon(\tau-\tau^\prime)}}{(\epsilon^2-\om_1^2+i0)}
=\delta^{ab}\frac{1}{2\om_1} \e^{-\om_1 |\tau-\tau^\p|}
{}~~~~~(\hbox{for }\l=0)\,.
\label{free}
\ee

The equations~\rf{eq2}, \rf{eqff} and \rf{u} can be finally rewritten in
the mixed representation as follows:
\be
\left\{\frac{d^2}{d\tau^2} -\om^2- v(\tau)  \right\}
\frac 1{N_1} D_{\c}^{bb}(\om ;\tau,\tau^\p)= -\delta(\tau-\tau^\p) ,
\label{feq2}
\ee
\be
\left\{ \frac{d^2}{d\tau^2} -\om^2- v(\tau)  \right\}
\frac 1{N_2} D_{\f}^{\b\b}(\om ;\tau,\tau^\p)= -\delta(\tau-\tau^\p)
\label{feq3}
\ee
and
\bea
v(\tau) = \l (\Phi^2(\tau) +\Psi^2(\tau)) +
\frac{\l}{2\pi^2} \left( \int_{m_1}d\om \;\om \sqrt{\om^2-m_1^2}
D_{\c}^{bb}(\om ;\tau,\tau) +\right. \non
+\left. \int_{m_2}d\om \;\om \sqrt{\om^2-m_2^2}
D_{\f}^{\b\b}(\om ;\tau,\tau) \right) \,.~~~~~~~~~~~
\label{fu}
\eea

To understand the meaning of the summed amplitudes
$D_{\c}^{bb}(\om ;\tau,\tau^\p)$ and $D_{\f}^{\b\b}(\om ;\tau,\tau^\p) ,$
let us note that
$D_{\c}^{bc}(\om ;\tau,\tau^\p)$ has in the index space the structure
\be
D_{\c}^{bc}(\om ;\tau,\tau^\p) =\left(\delta^{bc}
-\frac{\xi_1^b\xi_1^c}{\xi_1^2}\right)
G^T_{\c}(\om ;\tau,\tau^\p)+ \frac{\xi_1^b\xi_1^c}{\xi_1^2}
G^{S}_{\c}(\om ;\tau,\tau^\p)\,.
\label{structure}
\ee
The amplitudes $G^T$ and $G^{S}$ are associated, respectively, with
the {\it tensor\/} and {\it singlet\/}
$O(N_1)$-states of the two incoming particles.
The averaged over the $O(N_1)$-indices of two
incoming particles amplitude is
\be
\frac{1}{N_1} D_{\c}^{bb}(\om ;\tau,\tau^\p) =
\left(1- \frac{1}{N_1}\right)G^T_{\c}(\om ;\tau,\tau^\p) +
\frac 1{N_1} G^{S}_\c(\om ;\tau,\tau^\p)
\label{GTS}
\ee
while the generating function for the symmetrized  over all the $n$+$2$
$O(N_1)$-indices amplitudes is given by $G^{S}$:
\be
D^{ab}_\c (\om ;\tau,\tau^\p) \frac{\xi_1^a \xi_1^b}{\xi_1^2}
= G^{S}_\c (\om ;\tau,\tau^\p)\,.
\ee
One sees that the averaged amplitude which enters \eq{fu}
coincides with $G^T_\chi$  at large $N_1$.

We can now rewrite Eqs.~\rf{feq2} and \rf{feq3} at large $N$ as
the following equations for $G^T_{\c}(\om ;\tau,\tau^\p)$ and
$G^T_{\f}(\om ;\tau,\tau^\p)$ (the latter is defined similarly to
Eq.~\rf{GTS}:
\be
\left\{\frac{d^2}{d\tau^2} -\om^2- v(\tau) \right\}
G^T_\c(\om ;\tau,\tau^\p)=
-\delta(\tau-\tau^\p) \,,
\label{eqforG}
\ee
\be
\left\{\frac{d^2}{d\tau^2} -\om^2- v(\tau) \right\}
G^T_{\f} (\om ;\tau,\tau^\p)=
-\delta(\tau-\tau^\p) \,,
\label{eqforF}
\ee
while $v(\tau)$ is related to $G^T_\chi(\om;\tau,\tau)$
and $G^T_\phi(\om;\tau,\tau)$ by
\bea
v(\tau) = \l (\Phi^2(\tau)+\Psi^2(\tau)) +
\frac{\l N_1}{2\pi^2}\int_{m_1}d\om \;\om \sqrt{\om^2-m_1^2} \;
G^T_{\c} (\om ;\tau,\tau)+ \non
+\frac{\l N_2}{2\pi^2}\int_{m_2}d\om \;\om \sqrt{\om^2-m_2^2} \;
G^T_{\f} (\om ;\tau,\tau) \,.~~~~~~
\label{uvsG}
\eea
Therefore, only the tensor amplitudes enter at large $N_{1,2}$.

The transformation
from  Eqs.~\rf{recurrencea}, ~\rf{recurrenceb}, \rf{recurrenceD}
and \rf{recurrenceG} into
Eqs.~\rf{eq1}, \rf{eqf}, \rf{eqforG}, \rf{eqforF} and \rf{uvsG} is
similar to the one of Refs.~\cite{AKP93a,Vol93a,Smi93a,AKP93c,Mak94}.
These equations form the closed set
which will be solved in the next section.

\subsection{The operator derivation}

The structure of
Eqs.~\rf{eq1}, \rf{eqf}, \rf{eqforG}, \rf{eqforF} and \rf{uvsG} is
easy to understand through the use of the functional
technique~\cite{Bro92}
which relates $\tau$ to the time variable $t\equiv x_0$ of
the fields $\chi^a(x)$ and $\phi^\alpha(x)$
by
\be
\tau = it \,.
\ee
Using the definitions~\rf{defA} and \rf{defB}, one  rewrites
\rf{Phi} and \rf{Psi} as
\bea
\Phi^a(\tau) &=& m_1 \xi_1^a \e^{m_1\tau} \non & &+ \sum_{n,k\geq3}
\frac{1}{n!k!} \LA b_1\ldots b_n,\b_1 \ldots \b_k |\chi^a(0) |0 \RA
\xi_1^{b_1} \cdots \xi_1^{b_n} \xi_2^{\b_1} \cdots \xi_2^{\b_k}
\e^{(nm_1+km_2)\tau} \non
&=& \sum_{n,k\geq1}
\frac{1}{n!k!} \LA b_1\ldots b_n,\b_1 \ldots \b_k |
\chi^a(\vec{0},-i\tau) |0 \RA
\xi_1^{b_1} \cdots \xi_1^{b_n} \xi_2^{\b_1} \cdots \xi_2^{\b_k}
\label{PhiA}
\eea
and
\bea
\Psi^{\a}(\tau) &= &m_2 \xi_2^{\a} \e^{m_2\tau} \non & &+ \sum_{n,k\geq3}
\frac{1}{n!k!} \LA b_1\ldots b_n,\b_1 \ldots \b_k | \phi^\a (0) |0 \RA
\xi_1^{b_1} \cdots \xi_1^{b_n} \xi_2^{\b_1} \cdots \xi_2^{\b_k}
\e^{(nm_1+km_2)\tau} \non
&=& \sum_{n,k\geq1}
\frac{1}{n!k!} \LA b_1\ldots b_n,\b_1 \ldots \b_k |
\phi^\a (\vec{0},-i\tau) |0 \RA
\xi_1^{b_1} \cdots \xi_1^{b_n} \xi_2^{\b_1} \cdots \xi_2^{\b_k}
\,.
\label{PsiB}
\eea
Similarly, using \eq{D} and \rf{Dphi}, one sees
that $v(\tau)$ which is defined by \eq{u} is nothing but the sum of
the matrix elements:
\bea
v(\tau) = \hspace{12cm} \non \l \sum_{n,k=0}^\infty \frac{1}{n!k!}
\LA  b_1\ldots b_n ,\b_1\ldots \b_k |
:\!(\chi^2(0) + \phi^2(0))\!: |0 \RA
\xi_1^{b_1} \cdots \xi_1^{b_n} \xi_2^{\b_1} \cdots \xi_2^{\b_k}
\e^{(nm_1 +km_2)\tau} \non
 =  \l \sum_{n,k=0}^\infty \frac{1}{n!k!}
\LA  b_1\ldots b_n ,\b_1\ldots \b_k |
:\!(\chi^2(\vec{0},-i\tau) + \phi^2(\vec{0},-i\tau))\!: |0 \RA
\xi_1^{b_1} \cdots \xi_1^{b_n}
\xi_2^{\b_1} \cdots \xi_2^{\b_k}\,,
\label{ume}
\eea
where $:...:$ stands for a normal odering.

To simplify the notations, let us introduce the coherent state
\be
\LA  \Xi | \right. ~~=
\sum_{n,k=0}^\infty \frac{1}{n!k!}
\xi_1^{b_1} \cdots \xi_1^{b_n}
\xi_2^{\b_1} \cdots \xi_2^{\b_k}\,
\LA  b_1\ldots b_n ,\b_1\ldots \b_k | \right. ~~\,.
\label{coherent}
\ee
Then Eqs.~\rf{PhiA}, \rf{PsiB} and \rf{ume} can be written as follows:
\be
\Phi^a(\tau) = \LA \Xi | \chi^a(\vec{0},-i\tau) |0 \RA \,,
\label{PhiAX}
\ee
\be
\Psi^\a(\tau) = \LA \Xi | \phi^\a(\vec{0},-i\tau) |0 \RA
\label{PsiBX}
\ee
and
\be
v(\tau) = \LA \Xi |:\! (\chi^2(\vec{0},-i\tau)+\phi^2(\vec{0},-i\tau))\!:
|0\RA\,.
\label{umeX}
\ee

Eqs.~\rf{eq1}, \rf{eqf}, \rf{eqforG}, \rf{eqforF} and \rf{uvsG} can
alternatively be deduced  directly from
the \op \ equations for the quantum fields $\chi^a$ and $\phi^{\a}$:
\be
(-\partial_{\mu} \partial_{\mu} -m_1^2)\chi^a -
\l :\!(\chi^2 + \phi^2) \chi^a \!:\, =0 ,
\label{op1}
\ee
and
\be
(-\partial_{\mu} \partial_{\mu} -m_2^2)\phi^{\a}  -
\l :\!(\chi^2 + \phi^2) \phi^{\a} \!:\,=0 .
\label{op2}
\ee

Let us take the matrix elements of Eqs.~\rf{op1} and \rf{op2}
between the states $\LA \Xi| \right.$ and $\left. |0\RA $. We get
\bea
\left\{ \frac{\partial^2}{\partial \tau^2}+
\frac{\partial ^2}{\partial \vec{x}^2} -m_1^2\right\}
\LA \Xi |\chi^a(\vec{x},-i\tau) |0 \RA \non -
\l \LA \Xi| :\!(\chi^2(\vec{x},-i\tau)  +
\phi^2(\vec{x},-i\tau) ) \chi^a(\vec{x},-i\tau) \!:| 0 \RA =0 ,
\label{op1X}
\eea
and
\bea
\left\{ \frac{\partial^2}{\partial \tau^2}+
\frac{\partial ^2}{\partial \vec{x}^2} -m_2^2\right\}
\LA \Xi |\phi^\a(\vec{x},-i\tau) |0 \RA \non -
\l \LA \Xi |:\!(\chi^2(\vec{x},-i\tau)  +
\phi^2(\vec{x},-i\tau) ) \phi^\a(\vec{x},-i\tau) \!:| 0 \RA =0 \,.
\label{op2X}
\eea
Due to translational invariance one has
\be
\LA \Xi | \chi^a(\vec{x},-i\tau) |0 \RA = \Phi^a(\tau) \,,
\label{PhiAXx}
\ee
\be
\LA \Xi | \phi^\a(\vec{x},-i\tau) |0 \RA =\Psi^\a(\tau)
\label{PsiBXx}
\ee
so that the partial derivative w.r.t.\ $x$ vanishes.

In the large $N$ limit one can split the matrix elements of
the \op s
$:\!(\chi^2 + \phi^2) \chi^a :\!$ and $:\!(\chi^2 + \phi^2) \phi^{\a} \!:$ as
follows
$$
\LA a_1\ldots a_n ,\b_1 \ldots \b_k |
:\!(\chi^2 + \phi^2) \chi^a \!: |0\RA =
$$
$$=\sum_{p,p'} \sum_{n_1 +n_2 =n, \atop k_1 +k_2 =k} \fb
\LA p(a_1)\ldots p(a_{n_1}) ,p'(\b_1) \ldots p'(\b_{k_1}) |
:\!(\chi^2 + \phi^2)\!: |0\RA \times $$
\be
\times  \LA p(a_{n_1 + 1})\ldots p(a_n) ,p'(\b_{k_1 +1}))
\ldots p'(\b_k) |\chi^a |0\RA  ,
\label{factorization}
\ee
where $p,$ $p'$ stand for permutations.
This formula holds for $n$ and $k$ finite as $N\ra\infty$ and
extends  the standard {\it factorization\/} of $O(N)$-singlet operators
at large $N$  (see e.g.\ Ref.~\cite{Mak83}) to the case
of multiparticle production.

Using the definitions~\rf{coherent},
\rf{PhiAX}, \rf{PsiBX} and \rf{umeX}, we rewrite \eq{factorization}
in the form
\bea
\LA \Xi| :\!(\chi^2(\vec{x},-i\tau)  +
\phi^2(\vec{x},-i\tau) ) \chi^a(\vec{x},-i\tau) \!:| 0 \RA \non
=\LA \Xi |:\!(\chi^2(\vec{x},-i\tau) +\phi^2(\vec{x},-i\tau) )\!: | 0 \RA
\LA \Xi | \chi^a(\vec{x},-i\tau) | 0 \RA = v(\tau) \Phi^a(\tau)
\label{fact1}
\eea
and analogously
\bea
\LA \Xi| :\!(\chi^2(\vec{x},-i\tau)  +
\phi^2(\vec{x},-i\tau) ) \phi^\a(\vec{x},-i\tau) \!:| 0 \RA \non
=\LA \Xi |:\!(\chi^2(\vec{x},-i\tau) +\phi^2(\vec{x},-i\tau) )\!: | 0 \RA
\LA \Xi | \phi^\a(\vec{x},-i\tau) | 0 \RA = v(\tau) \Psi^\a(\tau)
\label{fact2}
\eea
where translational invariance of the averages, which results in
\be
\LA \Xi| :\!(\chi^2(\vec{x},-i\tau)  +
\phi^2(\vec{x},-i\tau) ) \!:| 0 \RA = v(\tau) \,,
\label{umeXx}
\ee
and Eqs.~\rf{PhiAXx}, \rf{PsiBXx} have been used.
Substituting \rf{fact1} and \rf{fact2} into Eqs.~\rf{op1X} and
\rf{op2X}, we obtain
Eqs.~\rf{eq1} and \rf{eqf}.

To infer the remaining Eqs.~\rf{eqforG}, \rf{eqforF} and \rf{uvsG},
let us represent $D^{ab}_\chi(\om_1;\tau,\tau^\p)$
and $D^{\a\b}_\phi(\om_2;\tau,\tau^\p)$
given by~\rf{Fourierc} and \rf{Fourierf} as
\bea
D^{ab}_\chi(\om_1;\tau,\tau^\p) = \int \frac{d^3 \vec{x}}{(2\pi)^3}
\e^{i\vec{p}\vec{x}}
\LA \Xi | \hbox{T} \chi^a(\vec{x},-i\tau) \chi^b(\vec{0},-i\tau^\p)| 0 \RA
\non - \delta^{(3)}(\vec{p}) \Phi^a(\tau) \Phi^b(\tau^\p)
\label{DDchi}
\eea
and
\bea
D^{\a\b}_\phi(\om_1;\tau,\tau^\p) = \int \frac{d^3 \vec{x}}{(2\pi)^3}
\e^{i\vec{p}\vec{x}}
\LA \Xi | \hbox{T} \phi^\a(\vec{x},-i\tau) \phi^\b(\vec{0},-i\tau^\p)| 0 \RA
\non - \delta^{(3)}(\vec{p}) \Psi^\a(\tau) \Psi^\b(\tau^\p) \,.
\label{DDphi}
\eea
Here the subtractions cancel disconnected parts
which are not included in the definition of $D^{ab}_\chi$ and $D^{\a\b}_\phi$
and $\hbox{T}$ stands for the T-product.
Notice that \eq{uvsG} can immediately be rewritten in the form \rf{umeX}
using ~\rf{DDchi} and \rf{DDphi}.
The terms involving the propagators $D^{aa}$ and $D^{\a\a}$
in the equation for $v(\tau)$ are associated with the connected part of
the \op \ $:\!\chi^2 + \phi^2\!:$.

The operator equations~\rf{op1} and \rf{op2} result in
the following equations for the T-products:
\bea
\left\{ \frac{\partial^2}{\partial \tau^2}+
\frac{\partial ^2}{\partial \vec{x}^2} -m_1^2\right\}
\LA \Xi | \hbox{T} \chi^a(\vec{x},-i\tau)\chi^b(\vec{0},-i\tau^\p) |0 \RA
\non -
\l \LA \Xi| \hbox{T} (\chi^2(\vec{x},-i\tau)  +
\phi^2(\vec{x},-i\tau) ) \chi^a(\vec{x},-i\tau)
\chi^b(\vec{0},-i\tau^\p)| 0 \RA
\non
= -\delta^{ab} \delta(\tau-\tau^\p)
\delta^{(3)}(\vec{x})
\label{sd1X}
\eea
and
\bea
\left\{ \frac{\partial^2}{\partial \tau^2}+
\frac{\partial ^2}{\partial \vec{x}^2} -m_2^2\right\}
\LA \Xi | \hbox{T} \phi^\a(\vec{x},-i\tau)\phi^\b(\vec{0},-i\tau^\p) |0 \RA
\non -\l \LA \Xi| \hbox{T} (\chi^2(\vec{x},-i\tau)  +
\phi^2(\vec{x},-i\tau) ) \phi^\a(\vec{x},-i\tau)
\phi^\b(\vec{0},-i\tau^\p)| 0 \RA
\non
= -\delta^{\a\b} \delta(\tau-\tau^\p)
\delta^{(3)}(\vec{x}) \,,
\label{sd2X}
\eea
where the delta functions on the r.h.s.'s emerge in a standard way when
differentiating the T-product and using canonical commutation relations.

Eqs.~\rf{sd1X} and \rf{sd2X} are the lowest ones in the chain
of the Schwinger--Dyson equations which can be
extracted from the following identities
\be
\LA \Xi \left| \hbox{T} \frac{\delta S[\chi,\phi]}{\delta \chi^a(x)}
F[\chi,\phi] \right|0\RA = i
\LA \Xi \left| \hbox{T} \frac{\delta F[\chi,\phi]}{\delta \chi^a(x)}
\right|0\RA
\label{identity1}
\ee
and
\be
\LA \Xi \left| \hbox{T} \frac{\delta S[\chi,\phi]}{\delta \phi^\a(x)}
F[\chi,\phi] \right|0\RA = i
\LA \Xi\left|\hbox{T} \frac{\delta F[\chi,\phi]}{\delta \phi^\a(x)}
\right|0\RA
\label{identity2}
\ee
where $S$ is the action of the model and $F[\chi,\phi]$ is an
arbitrary functional of $\chi$ and $\phi$.
In particular, Eqs.~\rf{op1X} and \rf{op2X} are associated with $F=1$
while Eqs.~\rf{sd1X} and \rf{sd2X} correspond to
$F=\chi^b(\vec{0},-i\tau^\p)$ and $F=\phi^\b(\vec{0},-i\tau^\p)$,
respectively.
In the language of path integral
Eqs.~\rf{identity1} and \rf{identity2} result from the invariance of the
measure under an arbitrary variation of $\chi^a(x)$ or $\phi^\a(x)$.

Using the large-$N$ factorization
\bea
\LA \Xi| \hbox{T}(\chi^2(\vec{x},-i\tau)  +
\phi^2(\vec{x},-i\tau) ) \chi^a(\vec{x},-i\tau) \chi^b(\vec{0},-i\tau^\p)
| 0 \RA \non
=\LA \Xi |:\!(\chi^2(\vec{x},-i\tau) +\phi^2(\vec{x},-i\tau) )\! :| 0 \RA
\LA \Xi |\hbox{T} \chi^a(\vec{x},-i\tau)\chi^b(\vec{0},-i\tau^\p)  | 0 \RA
\label{fact1D}
\eea
and analogously
\bea
\LA \Xi| \hbox{T}(\chi^2(\vec{x},-i\tau)  +
\phi^2(\vec{x},-i\tau) ) \phi^\a(\vec{x},-i\tau) \phi^\b(\vec{0},-i\tau^\p)| 0
\RA \non =\LA \Xi |:\!(\chi^2(\vec{x},-i\tau) +\phi^2(\vec{x},-i\tau) )\!:
| 0 \RA
\LA \Xi | \hbox{T} \phi^\a(\vec{x},-i\tau)\phi^\b(\vec{0},-i\tau^\p) | 0 \RA
\,,
\label{fact2D}
\eea
as well as
Eqs.~\rf{umeXx}, \rf{eq1}, \rf{eqf} and upon integrating over $d^3 x$, we
obtain
Eqs.~\rf{eqforG}, \rf{eqforF} from Eqs.~\rf{sd1X}, \rf{sd2X}.

Some comments concerning the operator derivation
of Eqs.~\rf{eq1}, \rf{eqf}, \rf{eqforG}, \rf{eqforF} and \rf{uvsG}
are in order.
The factorization was crucial
to drop out the dependence on the spatial coordinates
in the interaction terms and, therefore, for the reduction
at large $N$ to the quantum mechanical problem
which is given by the closed set of
Eqs.~\rf{eq1}, \rf{eqf}, \rf{eqforG}, \rf{eqforF} and \rf{uvsG}
Its appearance is due to the special kinematics of the produced
particles whose wave functions do not depend on the spatial
coordinates. The factorization formula~\rf{factorization} holds
for $n$ and $k$ fixed as $N\ra\infty$ (notice that only $n_{1,2}\leq n$
and $k_{1,2}\leq k$ enter \eq{factorization}).
The corrections to the large-$N$ amplitudes are known~\cite{Vol93a,Smi93a}
to be controlled at large $n$ by the parameter $n^2 \l \sim n^2 /N$ (also by
$k^2 /N$ in the given case) and are due to rescatterings of the produced
particles.  Hence, equations
\rf{eq1}, \rf{eqf}, \rf{eqforG}, \rf{eqforF} and \rf{uvsG}
allow for the calculation of the multiparticle amplitudes only
for $n, k \ll \sqrt{N}$.

\newsection{The exact solution}

To solve the set of equations~\rf{eq1}, \rf{eqf},
\rf{eqforG}, \rf{eqforF} and \rf{uvsG}, let
us first look at Eqs.~\rf{eqforG} and \rf{eqforF} for a given $v(\tau)$.
These equations determine the Green functions
of the Schr\"odinger operator with the
potential $v(\tau)$ while $\tau$ plays the role of a $1$-dimensional
coordinate.
Equivalently $G^T(\om;\tau,\tau^\p)$ is the
matrix element of the resolvent
\be
G^T(\om ;\tau,\tau^\p) = \langle \tau |
\frac{1}{-D^2+\om^2+v}|\tau^\p\rangle
\ee
where $D$ stands for $d/d\tau$ for brevity.
One should then take the diagonal matrix element of the resolvent,
$G^T(\om;\tau,\tau)$, to substitute it into \eq{uvsG}
and determine $v(\tau)$ versus $\Phi^2$ and $\Psi^2 .$

The general solution of this problem for  arbitrary $v$
is given by the Gelfand--Diki\u{\i} formula~\cite{GD75}
\be
G^T(\om ;\tau,\tau) = R_\om[v] \equiv \sum_{l=0}^\infty
\frac{R_l[v]}{\om^{2l+1}}
\label{GD}
\ee
where the differential polynomials $R_l[v]$ are determined
recurrently by
\be
R_l[v] = \frac{1}{2^l} \left( \frac 12 D^2-v- D^{-1}vD\right)^l
\label{polynomial}
\cdot \frac 12
\ee
and the inverse operator is
\be
D^{-1} v(\tau) = \int^\tau_{-\infty} dx \;v(x) \,.
\ee
\eq{polynomial} stems from the fact that $R_\om[v]$ obeys
the third  order linear differential equation
\be
\fr 12 \left( \fr 12 D^3 -Dv -vD \right)R_\om[v] =
\om^2 DR_\om[v]\,.
\label{linear}
\ee
The polynomials $R_l[v]$ depend on $v$ and its derivatives
$v^{(s)}\equiv (D^s v)$.
The  first few polynomials are
\bea
R_0[v] = \frac 12\;,~~~ R_1[v] = - \frac v4\;,
{}~~~ R_2[v] = \frac {1}{16} (3v^2-v^{\p\p})\;, \non
 R_3[v] = -\frac {1}{64} (10v^3 -10vv^{\p\p} -5(v^\p)^2 +v^{(4)})\,, \ldots
\label{RRR}
\eea
while for  $\tau$-independent $v(\tau)=v_0$ one has
\be
R_\om[v_0] = \frac{1}{2\sqrt{\om^2+v_0}}
{}~~~~~\hbox{(} v_0=\hbox{const.)}
\label{constant}
\ee
which agrees with \eq{free} at $\tau=\tau^\p$.

Following Ref.~\cite{Mak94} we renormalize the above equations.
Let us introduce $v_0$ as
\be
v_0 = \frac{\l}{4\pi^2} \left( N_1 \int_{m_{1R}} d\om \sqrt{\om^2
-m_{1R}^2} +
N_2 \int_{m_{2R}} d\om \sqrt{\om^2
-m_{2R}^2} \right) ,
\ee
where the renormalized masses $m_{1,2R}$ are defined by the following
equations
\be
m_{1R}^2=m_{1}^2 + v_0 ,\;\;\; m_{2R}^2=m_2^2 + v_0 .
\label{renM}
\ee
We also introduce the renormalized coupling constant defined as
\be
\frac{1}{\l_R} = \frac{1}{\l} + \frac{N_1}{8\pi^2}
\int_{m_{1R}} d\om \frac{\sqrt{\om^2 -m_{1R}^2}}{\om^2} +
\frac{N_2}{8\pi^2}
\int_{m_{2R}} d\om \frac{\sqrt{\om^2 -m_{2R}^2}}{\om^2} .
\label{renL}
\ee
The renormalized potential $v_R (\tau)$ reads
$$
v_R (\tau) = v(\tau) -v_0=
$$
$$=\l_R (\Phi^2 +\Psi^2) + \frac{\l_R}{2\pi^2} \left[
N_1\int_{m_{1R}}d\om \;\om \sqrt{\om^2-m^2_{1R}}
\left(R_\om[v_R] -\frac{R_1[v_R]}{\om^3}-\frac{1}{2\om}\right) +\right. $$
\be
\left. + N_2 \int_{m_{2R}}d\om \;\om \sqrt{\om^2-m^2_{2R}}
\left(R_\om[v_R] -\frac{R_1[v_R]}{\om^3}-\frac{1}{2\om}\right) \right]\,.
\label{uvsGR}
\ee
Now everything is expressed in terms of
$m_{1,2R}$ and $v_R(\tau)$:
\be
\left\{D^2 - m_{1R}^2 -v_R(\tau) \right\} \Phi^a(\tau) =0\,,
\label{eq1R}
\ee
\be
\left\{D^2 - m_{2R}^2 -v_R(\tau) \right\} \Psi^a(\tau) =0\,,
\label{eq2R}
\ee
\be
\om = \sqrt{\vec{p}\;^2+m_{1,2R}^2} \,.
\ee
The integrals over $\om$ on the r.h.s.\ become convergent after
the renormalizations.
The meaning of renormalization is that one chooses the bare
quantities, $m^2_{1,2}$ and $\l$, to be dependent on the cut-off according
to Eqs.~\rf{renM} and \rf{renL} and render the renormalized ones,
$m^2_R$ and $\l_R$, cut-off-independent.

We can easily find an exact solution to these equations.
Let us denote
\be
L= \fr 12 \left( \fr 12 D^3 -Dv -vD \right)\,.
\ee
We shall look for $R_{\om} [v_R]$ in the following form
\be
R_{\om} [v_R] = a(\om) + x(\om) \Phi^2 + y(\om) \Psi^2 ,
\ee
where $a,$ $x$ and $y$ are constants which may depend on $\om .$
Using eq.~\rf{linear} and the fact~\cite{Mak94} that
$\Phi^2$ and $\Psi^2$ are the eigenvectors%
\footnote{This is
derived using solely Eqs.~\rf{eq1R} and \rf{eq2R} for an arbitrary
$v_R(\tau)$ and the fact that $\Phi^2(\tau)$
and $\Psi^2(\tau)$ vanish for $\tau\ra -\infty$.}
of the operator $D^{-1} L:$
\be
\fr{1}{2} \left( \fr 12 D^2-v_R- D^{-1}v_RD \right) \Phi^2 =
\fr{1}{2} D^{-1}\left( \fr 12 D^3-Dv_R- v_RD \right) \Phi^2
=D^{-1}Dm_{1R}^2 \Phi^2 =m_{1R}^2 \Phi^2 \, ,
\label{eigenvector}
\ee
\be
\fr{1}{2} \left( \fr 12 D^2-v_R- D^{-1}v_RD \right) \Psi^2 =
\fr{1}{2} D^{-1}\left( \fr 12 D^3-Dv_R- v_RD \right) \Psi^2
=D^{-1}Dm_{2R}^2 \Phi^2 =m_{2R}^2 \Psi^2 \, ,
\label{eigenvector1}
\ee
one gets
\be
LR_{\om} [v_R] = -\frac{a}{2} D v_R + xm_{1R}^2 D\Phi^2 + ym_{2R}^2
D\Psi^2 =
\ee
$$=\om^2 D (a+x\Phi^2 +y\Psi^2) .$$
We remove $D$ from both sides of the above equation and we get
\be
v_R = c + \frac{2x}{a} (m_{1R}^2 -\om^2) \Phi^2 +
\frac{2y}{a} (m_{2R}^2 -\om^2) \Psi^2 ,
\ee
where $c$ is a constant of integration.
Thus we have
\be
v_R = c + \alpha \Phi^2 + \b \Psi^2 ,
\ee
where the constants
\be
\a= - \frac{2x(\om^2 - m_{1R}^2)}{a} ,\;\;\;
\b= - \frac{2y(\om^2 - m_{2R}^2)}{a}
\ee
do not depend on $\om$ since $v_R$ does not.

We now substitute both expressions for $R_{\om}$ and $v_R$ into
the equation~\rf{uvsGR} which determines $v_R.$
One can easily see that $c= 0,$ $a=1/(2\om)$ (notice that this value of
$a$ agrees with the perturbation expansion) and
\be
\a = \frac{\l_R}{1+\frac{\l_R}{8\pi^2}I_1} ,\;\;\;
x =-\frac{\a}{4\om (\om^2 -m_{1R}^2)} ,
\label{alpha}
\ee
\be
\b = \frac{\l_R}{1+\frac{\l_R}{8\pi^2}I_2} ,\;\;\;
y =-\frac{\b}{4\om (\om^2 -m_{2R}^2)} ,
\label{beta}
\ee
where
\bea
I_1=N_1 \int_{m_{1R}}d\om \sqrt{\om^2 - m_{1R}^2}
\left[ \frac{1}{\om^2 - m_{1R}^2} - \frac{1}{\om^2} \right] + \non
+ N_2 \int_{m_{2R}} d\om \sqrt{\om^2 - m_{2R}^2}
\left[ \frac{1}{\om^2 - m_{1R}^2} - \frac{1}{\om^2} \right],
\eea
\bea
I_2=N_1 \int_{m_{1R}}d\om \sqrt{\om^2 - m_{1R}^2}
\left[ \frac{1}{\om^2 - m_{2R}^2} - \frac{1}{\om^2} \right] +
\non
+ N_2 \int_{m_{2R}} d\om \sqrt{\om^2 - m_{2R}^2}
\left[ \frac{1}{\om^2 - m_{2R}^2} - \frac{1}{\om^2} \right].
\eea
After some calculations one gets
\be
I_1= N_1 + N_2 \int_{m_{2R}}
d\om \sqrt{\om^2 -m_{2R}^2}
\left[ \frac{1}{\om^2 - m_{1R}^2} -\frac{1}{\om^2} \right]
\label{I1}
\ee
and
\be
I_2= N_2 + N_1 \int_{m_{1R}}
d\om \sqrt{\om^2 -m_{1R}^2} \left[ \frac{1}{\om^2 - m_{2R}^2}
-\frac{1}{\om^2} \right] .
\label{I2}
\ee
For $m_1=m_2$ these formulas recover the results~\cite{Mak94} for
the $O(N_1$$+$$N_2)$-case. The fact that the solution is unique can
be proven quite similarly to Ref.~\cite{Mak94}.

The diagonal resolvent reads
\be
R_{\om} [v_R] = \frac{1}{2\om} - \frac{\a\Phi^2}{4\om (\om^2 -m_{1R}^2)}
-\frac{\b\Psi^2}{4\om (\om^2 -m_{2R}^2)}.
\label{dres}
\ee
The appearance of the poles in the resolvent has two faces.

{\bf 1.} From one point of view it is not directly related
to the propagation of particles in the loops.
For example the pole in the resolvent
appears for the symmetric $O(N)$ case even at the tree level.
The pole is actually due to an existence of a strong classical external
field which induces a potential $v(\tau)$
and corresponds to a bound state of a quantum mechanical particle.
Thus the pole in the resolvent is actually a manifestation of a bound
state in the spectrum of the corresponding Schr\"odinger operator.
This is of course an obvious fact but it explains the appearance of an
extra pole when we switch on an additional particle.
Emissions of an additional particle induce a new term in the effective
potential which is {\it not\/} suppressed at large $N$ since we assume
that the wave function of an emitted particle is of order
$\sqrt{N} .$
This new potential is responsible for a new bound state of the quantum
mechanical particle.
This also explains why the residue of the additional pole in the
resolvent is not suppressed at large $N.$

A natural question would be as to
why the poles should be exactly at the
values of the masses of the particles ({\it a priori\/} these poles could be
anywhere).
The answer is simple: this is because we know that there exist
solutions~\cite{LRT93a} to the Schr\"odinger equation for $\omega=m_{1,2}$
which decrease at both infinities.
Thus we know that indeed they are the bound states with the energies
$m_{1,2} .$

{\bf 2.} On the other hand the
diagonal resolvent itself corresponds to a loop by
definition since it corresponds to propagator integrated over the energy
(\ie to the one with
coinciding ends in the $\tau$-space).
For simplicity let us consider the model of the $O(N_1)$-particle
with the mass $m_1$ and an
additional singlet one with the mass $m_2$, \ie $N_2=1$.
In perturbation theory the simplest correction to the propagator
of the $O(N_1)$-particle is given by a
diagram with a 4-vertex which produces 2 outgoing particles with the mass
$m_2 .$
To extract the diagonal resolvent we should make a loop of this propagator
which is done in the momentum space by
integrating  over the energy ($d \epsilon$) of the virtual particle.
Thus we easily see that the correction to the resolvent which is due to an
emission of two particles is proportional to
\be
R_\om^{(2)} \propto
\int^{\infty}_{-\infty} \frac{d\epsilon}{2\pi}
{{1}\over{(\omega^2 - \epsilon^2 -i0)
(\omega^2 - (\epsilon-2m_2)^2 -i0)}} ,
\ee
where $\omega^2= |p|^2 + m_1^2 ,$ and $\vec{p}$ is a spatial virtual momentum.
This integral is convergent and can be computed by closing the contour
down around $\epsilon= \omega-i0$ and $\epsilon =\omega- 2m_2 - i0 .$
It is easy to see that a pole appears at $\omega = m_2$:
\be
R_\om^{(2)} \propto
{{i}\over{4m_2 \omega (\omega - m_2 -i0)}} .
\label{R(2)}
\ee
This is exactly what is implied by \rf{dres}.

We thus have checked that there {\it should\/} be two poles in the resolvent
in the model of two particles with different masses.
These poles sit exactly at the masses of those particles.

We now can find the expressions for $\Phi^a$ and $\Psi^{\a} .$
We have the following equations
\be
\left\{D^2 - m_{1R}^2 - (\a \Phi^2 + \b \Psi^2)\right\} \Phi^a =0,
\ee
\be
\left\{D^2 - m_{2R}^2 - (\a \Phi^2 + \b \Psi^2)\right\} \Psi^{\a} =0.
\ee
These equations reduce to the classical ones
(with the renormalized coupling constant $\l_R$) if we renormalize
the fields $\Phi^a$ and $\Psi^{\a}$ as follows
\be
\tilde{\Phi}^a =\sqrt{\frac{\a}{\l_R}} \Phi^a =
\left(1 + \frac{\l_R}{8\pi^2} I_1 \right)^{-1/2} \Phi^a ,\;\;\;
\tilde{\Psi}^{\a} =\sqrt{\frac{\b}{\l_R}} \Psi^{\a} =
\left(1 + \frac{\l_R}{8\pi^2} I_2 \right)^{-1/2} \Psi^{\a}.
\ee

The solution to these equations is given by the expressions
which are a straightforward generalization of those found
in Ref.~\cite{LRT93a}:
\be
\tilde{\Phi}^a =
\tilde{z}_1^a \left( 1-2\l_R \frac{k}{m_{2R}^2} \tilde{z}^2_2 \right)
\left( 1- \frac{2\l_R}{m_{1R}^2} \tilde{z}^2_1
-\frac{2\l_R}{m_{2R}^2} \tilde{z}^2_2
+ \l_R^2 \frac{k^2}{m^2_{1R} m^2_{2R}} \tilde{z}_1^2 \tilde{z}_2^2
\right)^{-1} ,
\label{rubC}
\ee
\be
\tilde{\Psi}^{\b} =
\tilde{z}_2^{\b} \left( 1+2\l_R \frac{k}{m_{1R}^2} \tilde{z}^2_1 \right)
\left( 1- \frac{2\l_R}{m_{1R}^2} \tilde{z}^2_1
-\frac{2\l_R}{m_{2R}^2} \tilde{z}^2_2
+ \l_R^2 \frac{k^2}{m^2_{1R} m^2_{2R}} \tilde{z}_1^2 \tilde{z}_2^2
\right)^{-1} ,
\label{rubF}
\ee
where
\be
k=\frac{m_{1R}-m_{2R}}{m_{1R}+m_{2R}} ,
\ee
and
\be
\tilde{z}_1^a = \xi_1^a \sqrt{\frac{\a}{\l_R}} e^{m_{1R} \tau}=
\xi_1^a e^{m_{1R} \tau} \left(1 + \frac{\l_R}{8\pi^2} I_1 \right)^{-1/2}
,
\ee
\be
\tilde{z}_2^{\a} = \xi_2^{\a} \sqrt{\frac{\b}{\l_R}} e^{m_{2R} \tau} =
\xi_2^{\a} e^{m_{2R} \tau} \left(1 + \frac{\l_R}{8\pi^2} I_2 \right)^{-1/2}
{}.
\ee

\begin{figure}[tbp]
\unitlength=1.00mm
\linethickness{0.6pt}
\centering
\begin{picture}(118.00,38.00)(11,90)
\put(70.00,114.00){\circle{14.00}}
\multiput(77.50,114.00)(7.0,3.50){3}{
    \line(2,1){4.00}}
\multiput(77.50,114.00)(7.0,-3.50){3}{
    \line(2,-1){4.00}}
\put(61.00,114.00){\line(-2,-1){18.00}}
\put(61.00,114.00){\line(-2,1){18.00}}

\end{picture}
\caption[x]   {\hspace{0.2cm}\parbox[t]{13cm}
{\small
   The one-loop diagram for the process $\chi+\chi \ra \phi+\phi$
   which possesses an imaginary part for $m_{2R}>m_{1R}$.
   given by \eq{IMG}. }}
\label{fig5}
\end{figure}
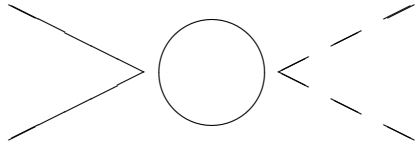

Notice that the constants $\a$ and $\b$ are in general complex.
By assuming for definiteness that $m_{2R}^2 \geq m_{1R}^2$ one can easily see
that the constant $\a$ is real while the imaginary part of $\b$ is given
by the following equation
\be
{\rm Im} \frac{1}{\b} = -\frac{N_1}{8\pi} \cdot \frac{\sqrt{m_{2R}^2
-m_{1R}^2}}{2m_{2R}} .
\label{IMG}
\ee
The appearance of this imaginary part
is in contrast to the $O(N)$ model with unbroken symmetry where
the amplitudes are real~\cite{Mak94}.
In the present model the imaginary part appears in $\a$ or $\b$
because the masses of $O(N_1)$ and $O(N_2)$ particles are not equal.
Indeed in a particular example $m_{2R}^2 \geq m_{1R}^2$
an imaginary part appears in the
amplitude of scattering of two $\chi$ particles  due to a possibility of
an inelastic process $\chi +\chi \to \phi +\phi .$
This imaginary part is associated with the diagram of Fig.~\ref{fig5}.
A direct calculation of the imaginary part of the diagram of Fig.~\ref{fig5}
gives the same expression~\rf{IMG}.

In the case of $N_2 =1$ which is discussed above, we drop out the contributions
proportional to $D_{\f}^{\a\a}$ in the recurrence equations.
Therefore this case can be easily recovered from Eqs.~\rf{rubC} and
\rf{rubF} by taking formally $N_2 =0$ in
Eqs.~\rf{I1} and \rf{I2} for $I_1$ and $I_2$ which enter the
expressions~\rf{alpha} and \rf{beta} for the constants $\a$ and $\b .$
The correct imaginary part immediately appears
for $m_2\geq m_1$ after integrating~\rf{R(2)}
over $\omega .$
This imaginary part corresponds to on-mass-shell virtual $O(N_1)$
particles emitting heavier singlet particles (the wave functions for incoming
particles are factorized out from this diagram and therefore it can be
thought as a contribution for the forward scattering for singlet
particles).

A generalization of the equations for the
generating functions $\Phi ,$ $\Psi$ to the case of $O(N_1) \times \ldots
\times O(N_s)$ scalar field is straightforward ($s$ is a positive
integer).
In such a general case which is described by the Lagrangian
\be
{\cal L} = \frac 12 \sum_{i=1}^s (\partial_\mu \phi_i) (\partial_\mu \phi_i)-
\frac 12 \sum_{i=1}^s m_i^2 \phi^2_i - \frac\l4 \left( \sum^s_{i=1} \phi_i^2
\right)^2 \,,
\ee
the equations for the generating functions for the amplitudes
of the multiparticle production at the threshold by a single particle
are given by equations similar to the classical ones while
the effect of loops in the large $N$ is reduced to a renormalization
of the masses and the coupling constant and to a (complex) renormalization
of the wave functions $\xi_i$ of the external particles.

The potential of the associated quantum mechanical problem is
\be
v_R = \sum_i \a_i \Phi_i^2 ,
\ee
where $\Phi_i^a$ are the generating functions for the amplitudes of
multiparticle production by a single particle $\phi_i^a ,$
and the parameters $\a_i$ are defined as follows
\be
\frac{1}{\a_i} = \frac{1}{\l_R} + \frac{1}{8\pi^2} \sum_j
N_j \int_{m_{jR}} d\om \sqrt{\om^2 - m_{jR}^2}
\left[ \frac{1}{\om^2 - m_{iR}^2} -\frac{1}{\om^2}\right] .
\ee
Here $m_{iR}$ stand for a renormalized mass of the $i$-particle.
The diagonal resolvent reads
\be
R_{\om} [v_R] = \frac{1}{2\om} - \sum_i \frac{\a_i \Phi_i^2}{4\om (\om^2
- m_{iR}^2)} .
\ee
This resolvent corresponds to a nullification of all amplitudes
except for those associated with the production of two identical particles.

\newsection{A model with spontaneously broken symmetry}

\subsection{The Schwinger--Dyson equations}

Let us now consider the model of an $O(N_1)$ scalar field and a singlet
scalar field with the wrong sign of the mass square.
The Lagrangian reads
\be
{\cal L} = \fr 12 (\partial_\mu \chi^b) (\partial_\mu \chi^b)+
\fr 12 (\partial_\mu \phi) (\partial_\mu \phi)-
\fr 12 m_1^2({\chi}^b{\chi}^b) -\fr 12 m_2^2(\phi^2)
-\fr 14 \l (\phi \phi + {\chi}^b{\chi}^b)^2 ,
\label{lagrangiansb}
\ee
where $m_{2}^2<0$ and $m_1^2 \geq -|m_2^2|$ can be either positive or
negative \footnote{Notice that in this case the vacuum state corresponds
to an unbroken $O(N_1)$ symmetry.}.
The interaction term has an $O(N)$-symmetry with $N=N_1$$+$$1$
which is explicitly broken for $m_1^2 \neq m_2^2$ by the mass term.
The Lagrangian~\rf{lagrangiansb} is invariant for $m_1^2 \neq m_2^2$
under the reflection: $\phi \ra -\phi$.

At the classical level one can see that the reflection symmetry
is spontaneously broken and the field $\f$ acquires a vacuum
expectation value
\be
\f_0=\frac{|m_2|}{\sqrt{\l}}
\ee
while the masses of the physical particles are
\be
m_\chi = \sqrt{m_1^2 +|m_2^2|}\,,~~~~~~m_\phi=\sqrt{2} |m_2|
{}~~~~~~~~\hbox{(tree level)}\,.
\label{masses}
\ee
If the $O(N)$-symmetry were not broken explicitly by the mass term
(\ie if $m_1=m_2$), one would get $N-1$ Goldstone bosons with the mass
$m_\chi=0$.

The classical equations of motion read
\bea
\{\d_{\mu}^2 + m_1^2 + \l (\c^a \c^a + \f^2)\} \c^b =0\,, \non
\{\d_{\mu}^2 - |m_2^2| + \l (\c^a \c^a + \f^2)\} \f =0 \,.
\label{cleq}
\eea

At the classical level one can shift the field $\f = \f' +\f_0 .$
Then the classical equations of motion take the form
\bea
\{\d_{\mu}^2 + (m_1^2 +|m_2^2|) +2\l\f_0 \f' + \l (\c^a \c^a + \f'^2)\} \c^b
=0\,, \non
\d_{\mu}^2\f' +2 |m_2^2|\f' +3\l \f_0 \f'^2
+ \l \f_0 \c^a\c^a +\l (\c^a \c^a + \f'^2) \f' =0.
\label{cleqsh}
\eea
One would expect that Eqs.~\rf{cleqsh} get simply generalized
to the quantum level through the substitution of
a normal ordering into \rf{cleqsh}.
However this is not quite correct because the vacuum expectation value
$\f_0$ can be renormalized due to the $\f' \c^2$ interaction.
Therefore it is convenient to consider the equations of motion for
the unshifted field $\f .$

One can extract the equations for the amplitudes to produce physical
particles by using the same method with the case
of no spontaneous symmetry breaking.
We shall assume that the field $\f$ contains a constant part which
corresponds to a correct (at the quantum level) vacuum expectation
value of $\f .$
We then take the matrix elements which are associated with
the production of physical particles at the threshold when the wave
functions of the produced particles are translationally invariant
similar to the procedure followed for the
case without spontaneous breaking of the $O(N)$ symmetry.

Indeed we have the following \op \ equations
\be
(\partial_{\mu} \partial_{\mu} +m_1^2)\chi^a +
\l :(\chi^2 + \phi^2) \chi^a : =0
\ee
and
\be
(\partial_{\mu} \partial_{\mu} -|m_2^2|)\phi +
\l :(\chi^2 + \phi^2) \phi :=0 ,
\ee
where $:\; :$ stands for a normal odering.
Let us take the matrix elements of these equations between the state
$\LA a_1\ldots a_n ,\f'^k| \right.$ and the vacuum where $\f'$ stands for a
physical particle that corresponds to the field $\f$.

One can see that
\bea
\Phi^a(\tau)
= \sum_{n,k\geq1}
\frac{1}{n!k!} \LA a_1\ldots a_n ,\f'^k |
\chi^a(\vec{0},-i\tau) |0 \RA
\xi_1^{b_1} \cdots \xi_1^{b_n} \xi_2^k
\label{PhiAsb}
\eea
and
\bea
\Psi(\tau)
=\sum_{n,k\geq1}
\frac{1}{n!k!} \LA a_1\ldots a_n ,\f'^k |
\phi (\vec{0},-i\tau) |0 \RA
\xi_1^{b_1} \cdots \xi_1^{b_n} \xi_2^k
\,.
\label{PsiBsb}
\eea
are nothing but the generating functions for the multiparticle production
from a single $\chi$ or $\phi'$ particle, respectively.
By introducing again the coherent state
\be
\LA \Xi' |  \right.~~=  \sum_{n,k\geq1}
\frac{1}{n!k!} \xi_1^{b_1} \cdots \xi_1^{b_n} \xi^k_2\,
\LA a_1\ldots a_n ,\f'^k |  \right.~~  \,,
\label{coherentsb}
\ee
we rewrite Eqs.~\rf{PhiAsb} and \rf{PsiBsb} as
\be
\Phi^a(\tau) = \LA \Xi' | \chi^a(\vec{0},-i\tau) |0 \RA \,,
\label{PhiAXsb}
\ee
\be
\Psi(\tau) = \LA \Xi' | \phi(\vec{0},-i\tau) |0 \RA \,.
\label{PsiBXsb}
\ee

In the large $N$ limit one can split the \op s
$:\!(\chi^2 + \phi^2) \chi^a\! :$ and $:\!(\chi^2 + \phi^2) \phi\!:$ as
follows
$$
\LA a_1\ldots a_n ,\f'^k |
:\!(\chi^2 + \phi^2) \chi^a\! : |0\RA =
$$
$$=\sum_p \sum_{n_1 +n_2 =n, \atop k_1 +k_2 =k} \fb
\LA p(a_1)\ldots p(a_{n_1}) , \f'^{k_1} |
:\! (\chi^2 + \phi^2)\!: |0 \RA \times $$
\be
\times  \LA p(a_{n_1 + 1})\ldots p(a_n) ,\f'^{k_2} |\chi^a |0\RA
\label{factorization1sb}
\ee
and
$$
\LA a_1\ldots a_n ,\f'^k |
:\!(\chi^2 + \phi^2) \f \!: |0 \RA =
$$
$$=\sum_p \sum_{n_1 +n_2 =n, \atop k_1 +k_2 =k} \fb
\LA p(a_1)\ldots p(a_{n_1}) , \f'^{k_1} |
:\!(\chi^2 + \phi^2)\!: |0 \RA \times $$
\be \times  \LA p(a_{n_1 + 1})\ldots p(a_n) ,\f'^{k_2} |\f |0 \RA
\label{factorization2sb}
\ee
where $p$ stand for permutations.
Using the definition~\rf{coherentsb}, Eqs.~\rf{factorization1sb} and
\rf{factorization2sb} can be rewritten in the simple form
\be
\LA \Xi'| :\!(\chi^2  +
\phi^2 ) \chi^a \!:| 0 \RA
=\LA \Xi'|:\!(\chi^2 +\phi^2 )\!: | 0 \RA
\LA \Xi' | \chi^a | 0 \RA
\label{fact1sb}
\ee
and
\be
\LA \Xi'| :\!(\chi^2  +
\phi^2) \phi \!:| 0 \RA
=\LA \Xi' |:\!(\chi^2+\phi^2 )\!: | 0 \RA
\LA \Xi' | \phi | 0 \RA \,.
\label{fact2sb}
\ee

The Schwinger--Dyson equations can now be derived quite similarly
to Subsect.~3.2. It is crucial again that one
drops out the dependence on 3D spatial coordinates
in the interaction terms.
This is correct due to the factorization of these
terms in the large-$N$ limit and the special kinematics of the produced
particles.
We thus get the following equations
\be
\left\{D^2 -m_\chi^2 -v(\tau)\right\}\c^a =0,
\ee
\be
\left\{D^2 - v(\tau)\right\}\f =0,
\ee
where
\be
v(\tau) = \l (\Phi^2 + \Psi^2 -\phi_0^2) + \l \int \frac{d^4 p}{(2\pi)^4}
D_{\c}^{bb} (\tau;p) .
\label{vsb}
\ee
Here
we have taken into account that $|m_2^2|=\l\phi_0^2$
and used the {\it physical\/} mass $m_{\c}$ of the field $\c$ which appears
when we take into account the non-vanishing vacuum expectation value of
the field $\f .$
At the classical level
$m_\chi$ is given by \eq{masses}.

The propagator $D_{\c}^{ab}$ obeys \eq{eq2} as for
the case without spontaneous breaking.
Introducing $D_{\c}^{ab} (\omega; \tau,\tau')$ (the mixed representation)
as in \eq{Fourierc} obeying
\be
\left\{D^2 -\omega^2 -v(\tau) \right\}
 \frac{1}{N} D_{\c}^{bb} (\omega;\tau,\tau') =
-\delta(\tau-\tau') ,
\ee
where $\om^2 = m_\chi^2 +|\vec{p}|^2$ \/ (${\vec{p}}$ stands for the
spatial components of momentum),
we can rewrite the potential $v(\tau)$ given by~\rf{vsb} as follows
\be
v(\tau) =\l (\Phi^2 + \Psi^2 -\phi_0^2) + \frac{\l N}{2\pi^2} \int_{m_{\c}}
d\omega \omega \sqrt{\omega^2 - m_{\c}^2} G^T (\omega;\tau,\tau) ,
\ee
where $G^T$ stands for the tensor part of $D_{\c}^{ab} .$
Actually these equations can be easily obtained
through the use of
generalized recurrence equations for the amplitudes to produce
particles $\c^a$ and $\f' .$
To do that we have to use the Lagrangian for the fields $\c$ and $\f'
=\f -\f_0$ which are physical at the classical level.
The only difference with the case of no spontaneous breaking is
a presence of the vertices $\c^2 \f'$ and $\f'^3 .$

We now renormalize the masses and the coupling constant.
Let us introduce
\be
v_0 =\frac{\l N}{4\pi^2} \int_{m_{\c}} d\omega \sqrt{\omega^2 -
m_{\c}^2} .
\ee
The renormalized masses are then defined as follows:
\be
m_{1R}^2 = m_1^2 + v_0 ,\;\;\; m_{2R}^2 = m_2^2 +v_0 ,
\ee
while for the renormalized coupling constant we have
\be
\frac{1}{\l_R} = \frac{1}{\l} + \frac{N}{8\pi^2} \int_{m_{\c}}
d\om \frac{\sqrt{\omega^2 -m_{\c}^2}}{\omega^2} .
\ee
We see that the physical mass of the field $\c$ is not renormalized,
$m_{\c}^2 =m_{1R}^2 - m_{2R}^2 =m_\chi^2.$

The renormalized potential is defined as
\be
v_R (\tau) = v(\tau) -v_0=
\label{vRsb}
\ee
$$=\l_R (\Phi^2 +\Psi^2 -\phi_{0}^2) + \frac{\l_R}{2\pi^2}
N_1\int_{m_{1R}}d\om \;\om \sqrt{\om^2-m^2_{1R}}
\left(R_\om[v_R] -\frac{R_1[v_R]}{\om^3}-\frac{1}{2\om}\right) .$$
Here $\phi_{0} =|m_{2R}|/\sqrt{\l_R}$ is the renormalized vacuum expectation
value of the field $\phi .$

Now everything is expressed in terms of
$m_{1,2R}$ and $v_R(\tau)$:
\be
\left\{D^2 - m_{\c}^2 -v_R(\tau) \right\} \Phi^a(\tau) =0\,,
\label{eqRB}
\ee
\be
\om = \sqrt{\vec{p}\;^2+m_{\c}^2} \,.
\ee
The integral over $\om$ on the r.h.s.\ of \eq{vRsb} becomes convergent after
the renormalizations.

\subsection{The exact solution}

We shall look for $R_{\om} [v_R]$ in the following form
\be
R_{\om} [v_R] = a(\om) + x(\om) \Phi^2 + y(\om) \Psi^2 ,
\ee
where $a,$ $x$ and $y$ are constants which may depend on $\om .$
Similarly with the previous case of no spontaneous breaking we shall use
Eq.~\rf{linear} and the fact that
\be
L\Phi^2 =\fr{1}{2} \left( \fr 12 D^3-D v_R- v_RD \right) \Phi^2 =
m_{\c}^2 D \Phi^2 \, ,
\label{eigenvectorBC}
\ee
\be
L \Psi^2 =\fr{1}{2} \left( \fr 12 D^3-D v_R-v_RD \right) \Psi^2 = 0.
\label{eigenvectorBP}
\ee

We obtain
\be
LR_{\om} [v_R] = -\frac{a}{2} D v_R + xm_{\c}^2 D\Phi^2 =
\ee
$$=\om^2 D (a+x\Phi^2 +y\Psi^2) .$$
By removing $D$ from both sides of the above equation we get
\be
v_R = c + \frac{2x}{a} (m_{\c}^2 -\om^2) \Phi^2 -
\frac{2y}{a} \om^2 \Psi^2 ,
\ee
where $c$ is a constant of integration.
Thus we have
\be
v_R = c + \a \Phi^2 + \b \Psi^2 ,
\ee
where the constants
\be
\a= - \frac{2x(\om^2 - m_{\c}^2)}{a} ,\;\;\;
\b= - \frac{2y\om^2}{a}
\ee
do not depend on $\om$ since $v_R$ does not.

We now substitute the expressions for $R_{\om}$ and $v_R$ into
the equation that defines $v_R.$
We have
\be
c+\a \Phi^2 +\b \Psi^2 = \l_R (\Phi^2 +\Psi^2 -\phi_0^2) +
\label{uravn}
\ee
$$+\frac{\l_R N}{2\pi^2} \int_{m_{\c}} d\om \om \sqrt{\om^2 -m_{\c}^2}
\left[ a - \frac{\a a}{2(\om^2 -m_{\c}^2)} \Phi^2
-\frac{\b a}{2 \om^2} \Psi^2 +\frac{1}{4\om^3} (c + \a \Phi^2 + \b
\Psi^2) -\frac{1}{2\om} \right] .$$
{}From the above equation we get
\be
\a = \l_R + \frac{\a \l_R N}{2\pi^2} \int_{m_{\c}} d\om \om
\sqrt{\om^2 -m_{\c}^2} \left[ \frac{1}{4\om^3} - \frac{a}{2(\om^2 -m_{\c}^2)}
 \right] ,
\label{renA}
\ee
\be
\b = \l_R + \frac{\b \l_R N}{2\pi^2} \int_{m_{\c}} d\om \om
\sqrt{\om^2 -m_{\c}^2} \left[\frac{1}{4\om^3} - \frac{a}{2\om^2} \right] .
\label{renB}
\ee
One can also extract an equation for the constant part of \eq{uravn}
\be
c + \b \phi_0^2 = \frac{\b \l_R N}{2\pi^2} \int_{m_{\c}} d\om \om
\sqrt{\om^2 -m_{\c}^2}
\left[ a -\frac{\b a}{2 \om^2} \phi_0^2 +\frac{1}{4\om^3} (c + \b
\phi_0^2) -\frac{1}{2\om} \right] .
\label{const}
\ee
We get three equations for four unknown parameters $\a ,$ $\b ,$ $a$ and
$c.$

To choose the correct solution to
Eqs.~\rf{renA}, \rf{renB} and \rf{const} we have to use
the following physical condition.
When the resolvent is expressed in terms of the physical (shifted) fields
$\Phi^a$ and $\Psi' =\Psi -\phi_0$ the constant part of the resolvent is
fixed in perturbation theory to be equal to $1/(2\om) .$
This means that
\be
a - \frac{a\b \phi_0^2}{2\om^2} = \frac{1}{2\om} ,
\ee
and hence
\be
a= \frac{\om}{2(\om^2 -\b\phi_0^2/2)} .
\ee
We substitute this expression into \eq{const} we get
\be
c=-\b \phi_0^2 .
\ee
For the parameters $\a$ and $\b$ we have
\be
\a = \frac{\l_R}{1 + \frac{\l_R N}{8\pi^2} J_1},
\ee
\be
\b = \frac{\l_R}{1 + \frac{\l_R N}{8\pi^2} J_2} ,
\ee
where
\be
J_1 = \int_{m_{\c}} d\om
\sqrt{\om^2 -m_{\c}^2} \left[ \frac{\om^2}{(\om^2 -
\b\phi_0^2/2)(\om^2 -m_{\c}^2)}
- \frac{1}{\om^2} \right] ,
\ee
\be
J_2 = \int_{m_{\c}} d\om
\sqrt{\om^2 -m_{\c}^2} \left[ \frac{1}{\om^2 -\b\phi_0^2/2}
- \frac{1}{\om^2} \right] .
\ee
It is easy to see that the parameters $\a$ and $\b$ are real in contrast
to the case without the spontaneous breaking of the $O(N)$-symmetry.

For the resolvent we get the following expression
\be
R_{\om} [v_R] = \frac{1}{2\om(\om^2 -\b\phi_0^2/2)} -
\frac{\a \om \Phi^2}{4(\om^2 - m_{\phi}^2/4) (\om^2-m_{\c}^2)}
-\frac{\b \Psi^2}{4\om(\om^2 - m_{\phi}^2/4)},
\label{resB}
\ee
where we identified the physical mass of the field $\Psi'$ with
\be
m_{\phi}^2 = 2\b \phi_0^2 .
\ee
One can see that the resolvent has two poles at $\om =m_{\c}$ and $\om =
m_{\phi}/2 .$
It can be easily checked in perturbation theory with respect to the
external lines at the tree level
for the resolvent that exactly these two poles appear to the lowest
orders.
Indeed the pole at $\om=m_{\c}$ corresponds to a diagram with two
external legs of the $\c$ field while the pole at $\om =m_{\f}/2$
comes from the diagram with a single $\f$ external leg.
It is worth noticing that by looking at \eq{resB} one might think that
there is a pole at $\om = m_{\f}/2$ in the amplitude to produce
two $\c$ particles which does not of course appear in perturbation
theory.
It is easy to check however that such a pole is cancelled if we take
into account the explicit solution for $\Phi^a$ and $\Psi$
which is given below.
One can also see that there are no poles at $\om =m_{\f}$ in the
resolvent.
This signals the fact that the amplitudes for the production of two external
$\f$ particles are cancelled in contrast to the case without
the spontaneous breaking.

Thus we get the following equations for the generating functions
\be
\left\{D^2 - m_{\c}^2 - (\a \Phi^2 + \b \Psi^2 -\b \phi_0^2)
\right\} \Phi^a =0,
\ee
\be
\left\{D^2 - (\a \Phi^2 + \b \Psi^2 -\b\phi_0^2)\right\} \Psi =0.
\ee
These equations reduce to the classical ones
(with the renormalized coupling constant $\l_R$) if we renormalize
the fields $\Phi^a$ and $\Psi$ as follows
\be
\tilde{\Phi}^a =\sqrt{\frac{\a}{\l_R}} \Phi^a =
\Phi^a \left(1 + \frac{\l_R N}{8\pi^2} J_1 \right)^{-1/2} ,\;\;\;
\tilde{\Psi} =\sqrt{\frac{\b}{\l_R}} \Psi =
\Psi \left(1 + \frac{\l_R N}{8\pi^2} J_2 \right)^{-1/2} .
\ee
The solution to the above equations which obeys the condition that at
$\l_R\to 0$
\be
\Phi^a \to m_{\c}
\xi_1^a e^{m_{\c} \tau} ,
\ee
$$\Psi \to \f_0 + m_{\f} \xi_2 e^{m_{\f} \tau}$$
is similar to that given in Ref.~\cite{LRT93a}:
\be
\tilde{\Psi} = \f_{0R} \left( 1 + \frac{\tilde{z}_2}{2\f_{0R}} +
\frac{8\l_R}{4m_{\c}^2 - m_{\f}^2} \tilde{z}_1^2 + \frac{4\l_R}{\f_{0R}}
\frac{2m_{\c} -m_{\f}}{(2m_{\c} + m_{\f})^3} \tilde{z}_2 \tilde{z}^2_1
\right) \times
\ee
$$\left( 1 - \frac{\tilde{z}_2}{2\f_{0R}} -
\frac{8\l_R}{4m_{\c}^2 - m_{\f}^2} \tilde{z}_1^2 + \frac{4\l_R}{\f_{0R}}
\frac{2m_{\c} -m_{\f}}{(2m_{\c} + m_{\f})^3} \tilde{z}_2 \tilde{z}^2_1
\right)^{-1} ,$$
\be
\tilde{\Phi}^a = \tilde{z}_1^a \left( 1-\frac{2m_{\c} -m_{\f}}{2m_{\c}
+m_{\f}} \frac{\tilde{z}_2}{2\f_{0R}}\right)
\left( 1 - \frac{\tilde{z_2}}{2\f_{0R}} -
\frac{8\l_R}{4m_{\c}^2 - m_{\f}^2} \tilde{z}_1^2 + \frac{4\l_R}{\f_{0R}}
\frac{2m_{\c} -m_{\f}}{(2m_{\c} + m_{\f})^3} \tilde{z}_2 \tilde{z}^2_1
\right)^{-1} .
\ee
Here
\be
\f_{0R} =
\phi_0 \sqrt{\frac{\b}{\l_R}}=
\f_0 \left(1 + \frac{\l_R N}{8\pi^2} J_2 \right)^{-1/2} ,
\ee
and
\be
\tilde{z}_1^a =m_{\c} \xi_1^a e^{m_{\c} \tau}
\left(1 + \frac{\l_R N}{8\pi^2} J_1 \right)^{-1/2} ,
\ee
\be
\tilde{z}_2 = m_{\f} \xi_2 e^{m_{\f}\tau}
\left(1 + \frac{\l_R N}{8\pi^2} J_2 \right)^{-1/2} .
\ee
Thus we finally get
\be
\Psi = \f_0 \left( 1 + \frac{\tilde{z_2}}{2\f_{0R}} +
\frac{8\l_R}{4m_{\c}^2 - m_{\f}^2} \tilde{z}_1^2 + \frac{4\l_R}{\f_{0R}}
\frac{2m_{\c} -m_{\f}}{(2m_{\c} + m_{\f})^3} \tilde{z}_2 \tilde{z}^2_1
\right) \times
\ee
$$\left( 1 - \frac{\tilde{z_2}}{2\f_{0R}} -
\frac{8\l_R}{4m_{\c}^2 - m_{\f}^2} \tilde{z}_1^2 + \frac{4\l_R}{\f_{0R}}
\frac{2m_{\c} -m_{\f}}{(2m_{\c} + m_{\f})^3} \tilde{z}_2 \tilde{z}^2_1
\right)^{-1} ,$$
\be
\Phi^a = z_1^a \left( 1-\frac{2m_{\c} -m_{\f}}{2m_{\c}
+m_{\f}} \frac{\tilde{z}_2}{2\f_{0R}}\right)
\left( 1 - \frac{\tilde{z_2}}{2\f_{0R}} -
\frac{8\l_R}{4m_{\c}^2 - m_{\f}^2} \tilde{z}_1^2 + \frac{4\l_R}{\f_{0R}}
\frac{2m_{\c} -m_{\f}}{(2m_{\c} + m_{\f})^3} \tilde{z}_2 \tilde{z}^2_1
\right)^{-1} .
\ee
We observe that this solution is real.

\newsection{Conclusions}

Let us briefly summarize our results.
We presented an exact solution in the large-$N$ limit
for the amplitudes of multiparticle
production at threshold in the $\phi^4$ theory with the $O(N)$-symmetry
which is softly broken to $O(N_1)\times \ldots\times O(N_s)$
($N$$=$$N_1$$+\ldots+$$N_s$) by the mass term.
We found that the effect of loops in this limit reduces to
a renormalization of coupling constants and masses.
Moreover as the masses $m_i$ of the $O(N_i)$-multiplets are
nonequal, the amplitudes of multiparticle production become
complex.
The explicit form of the solution demonstrates a nullification
of the multiparticle amplitudes with all legs on mass-shell.
The only non-vanishing amplitudes correspond to the processes $2$$\to$$2$
when the incoming (outgoing) particles are identical.

We also analyzed the model of the $O(N) + singlet$ scalar particles in the
case when the reflection symmetry is spontaneously broken.
In the limit of large $N$ the
effect of loops of the $O(N)$ field reduces to a renormalization
of the coupling constants and masses of the physical particles while no
imaginary part of amplitudes appears in this case.
Using the exact solution for this problem, we demonstrate a nullification
of amplitudes except for those with 2 incoming $O(N)$ particles and
2 outgoing $O(N)$ particles or 1 outgoing physical singlet particle.

Some problems remain unsolved.
As it has been demonstrated in Ref.~\cite{LRT93b},
the nullification of the tree level amplitudes for the model
of two scalar fields with further restrictions on kinematics,
when one particle, say $m_2$, can decay at rest into the others
with the mass $m_1$, is due to a symmetry
of the classical Lagrangian where the dependence of the fields
on the spatial 3D coordinate is discarded.
It is straightforward to check that such a symmetry exists at the tree
level in the $O(N_1)\times O(N_2)$ model as well.
Indeed the Lagrangian~\rf{lagrangian}
is invariant (up to terms which are total derivatives)
for the fields independent of the spatial coordinates
under the following transformations
\be
\delta\c^a = \epsilon \l \f^{\a} (\dot{\c}^a \f^{\a} -\dot{\f}^{\a}
\c^a) ,
\label{transformation}
\ee
$$\delta\f^{\a} = \epsilon [-\l \c^a(\dot{\c}^a \f^{\a} -\dot{\f}^{\a}
\c^a) +\frac{1}{2} (m_1^2 -m_2^2) \dot{\f}^{\a}] ,$$
where $\dot{\f}$ stands for $d \f/dt $
and $\epsilon$ is a constant parameter of the variation.
Following Ref.~\cite{LRT93b} one can see that the non-vanishing
amplitudes come from the resonance terms which appear in the
iteration procedure for solving the classical equations of motion
with the boundary conditions at $\l \to 0$:
\be
\c\to \xi_1^a e^{-im_1 t},\;\;\;
\f \to \xi_2^{\a} e^{-im_2 t} .
\ee

The explicit form of our solution shows that such a symmetry survives
at the multi-loop level (at large $N$) since the only effect of
multi-loop corrections reduces to a renormalization of the coupling
constants and masses.
Obviously we come to the same conclusion in the case of a spontaneously
broken reflection symmetry in the model of the $O(N)+ singlet$ scalar
particles. An interesting question about this symmetry
is whether it can be extended to arbitrary $m_1$ and $m_2$ when kinematics
requires nonvanishing spatial momenta of incoming particles.

As it is already mentioned in Subsect.~3.2, our results for the multiparticle
amplitudes are applicable when the number of produced particles is much less
than $\sqrt{N}$. They nicely show how the loop effects renormalize
tree amplitudes but do not take into account
diagrams with rescatterings in the final state which are expected to
restore unitarity. There exist at the moment two examples of such calculations:
1) a semiclassical calculation~\cite{GV93} for the $N=1$ case with a
spontaneously broken reflection symmetry, 2) models with any $N$ in $2+1$
dimensions which are solved~\cite{RS94} by summing infrared logarithms with the
aid of renormalization group.  It would be interesting to develop a technique
based on the Schwinger--Dyson equations which would be applicable for the case
 when the number of produced particles is of order $\sqrt{N}$.  We hope that
the operator formulation of the Schwinger--Dyson equations for the
multiparticle amplitudes which is proposed in this paper may be useful for this
purpose.

\subsection*{Acknowledgments}
M.A. and Yu.M. were sponsored, in part, by the Danish Natural Science
Research Council.
A.J. acknowledges the NBI high energy group for its hospitality.
His research was supported in part by a NATO grant GRG 930395.

\eop

\end{document}